\documentclass[a4paper]{article}
\usepackage{fullpage}
\usepackage{natbib}
\usepackage{graphicx}
\usepackage{graphics}
\usepackage{amssymb}
\usepackage{times}

\begin{document}
\author{P.~B. Carolan$^1$\thanks{e-mail: patrick.carolan@nuigalway.ie},
T. Khanzadyan$^1$, M. P. Redman$^1$, M. A. Thompson$^2$, P. A.
Jones$^{3,4}$, \\
M.R. Cunningham$^3$, R.M.~Loughnane$^1$, I. Bains$^5$ and E. Keto$^6$\\
\\
\small{$^1$Centre for Astronomy, School of Physics, National University of Ireland, Galway, University Road, Galway, Ireland}\\
\small{$^2$Centre for Astrophysics Research, University of Hertfordshire, College Lane, Hatfield, AL10 9AB, UK}\\
\small{$^3$School of Physics, University of New South Wales, Sydney, NSW 2052, Australia}\\
\small{$^4$Departamento de Astronom\'{i}a, Universidad de Chile, Casilla 36-D, Santiago, Chile}\\
\small{$^5$Centre for Astrophysics and Supercomputing, Swinburne University of Technology, PO Box 218, Hawthorn, VIC 3122, Australia}\\
\small{$^6$Harvard-Smithsonian Center for Astrophysics, 60 Garden
Street, Cambridge, MA 02138 }}
\date{}
\title{Supersonic turbulence in the cold massive core JCMT 18354-0649S\thanks{Based on observations collected at JCMT: The James Clerk Maxwell Telescope is operated by The Joint Astronomy Centre on behalf of the Science and Technology Facilities Council of the United Kingdom, the Netherlands Organisation for Scientific Research, and the National Research Council of Canada.}\\}
\maketitle
\begin{abstract}
An example of a cold massive core, JCMT 18354-0649S, a possible
high mass analogue to a low mass star forming core is studied.
Line and continuum observations from JCMT, Mopra Telescope and
\emph{Spitzer} are presented and modelled in detail using a 3D
molecular line radiative transfer code. In almost every way JCMT
18354-0649S is a scaled-up version of a typical low mass core with
similar temperatures, chemical abundances and densities. The
difference is that both the infall velocity and the turbulent
width of the line profiles are an order of magnitude larger. While
the higher infall velocity is expected due to the large mass of
JCMT 18354-0649S, we suggest that the dissipation of this highly
supersonic turbulence may lead to the creation of dense clumps of
gas that surround the high mass core.
\end{abstract}


\section{Introduction}

Studies of the formation of individual stars have traditionally
distinguished between low mass star formation and massive star
formation. Low mass stars tend to form in isolation in relatively
quiescent environments such as those seen in Taurus. In massive
star formation, such as in the Orion nebula, dense clusters of
thousands of low mass stars form alongside a few high mass stars
($> 8~{\rm M_\odot}$) which are capable of ionising the
surrounding remnant cloud. While the outcomes of these two modes
of star formation are vastly different, evidence is growing that
the physical conditions in the initial parent molecular clouds may
be very similar, differing only in size and mass scale between the
two paths
\citep{hillenbrand&hartmann98,patel-et-al05,fontani-et-al06}.

A key question that remains is how and when massive star formation
diverges from low mass star formation. In order to address this
problem, searches have been made for the high mass equivalents of
the starless or early protostellar cores seen in low mass star
formation \citep[see][for reviews]{bacmann00,andre04,bergtaf07}.
Infrared dark clouds, which are seen in absorption against 8
micron emission towards the galactic centre
\citep{egan98,perault96,teyssier02} are potential sites for
massive star forming clusters within which individual 'cold
massive cores' can be identified. Another strategy is to search
for cold massive cores amidst ongoing massive star formation
\citep{garay-et-al-04,hill-et-al-05,pillai07}. Three examples of
individual cold massive cores found via their dust continuum
emission are JCMT 18354-0649S \citep{wu.et.al05}, G333.125-0.562
\citep{lo-et-al07} and ISOSS J18364-0221 \citep{birkmann.et.al06}.
The continuum emission indicates that these sources are cold
($\sim 20~{\rm K}$) yet massive ($800~{\rm M_\odot}$, $1.8 \times
10^{3}~{\rm M_\odot}$ and $75~{\rm M_\odot}$ respectively) and
thus are potential future sites of massive star formation.

Different molecular lines trace different parts of a cloud and an
ideal tracer should faithfully trace the gas dynamics as deeply
into the cloud as possible. Candidate dynamical tracer species
include ${\rm N_2H^+}$, ${\rm HCO^+}$, HCN, ${\rm NH_3}$ and CO
amongst others. The interpretation of such lines is not trivial
even with a radiative transfer code since effects such as
depletion (when an atom/molecule freezes onto the surface of a
dust grain) and chemical variation coupled with dynamics such as
rotation, infall and outflow all contribute to the final emergent
line profile. Progress becomes possible however if several lines
from different species are used together.

\citet{wu.et.al05} observed JCMT\,18354-0649S (hereafter
JCMT\,18354S) in four species using the James Clark Maxwell
Telescope (JCMT): ${\rm H^{13}CO^+}$ $(J = 3~$--$~2)$,  ${\rm
C^{17}O}$ $(J = 2~$--$~1)$, ${\rm HCO^+}$ $(J = 3~$--$~2)$. They
found the molecular line profiles look similar in shape to those
of their low mass counterparts \citep[see also][]{fuller.et.al05}.
The main difference is in the line widths which are clearly
broader. The linewidth at half the intensity for the lines
C$^{18}$O (2 - 1), C$^{17}$O (2 - 1) and H$^{13}$CO$^{+}$ (3 - 2)
are 4.03, 3.78 and 3.24 respectively. Comparing those values with
the same parameter measured from the low mass core L483
\citep{carolan-et-al-08} we find the linewidth is 0.86, 1.09 and
0.66. This is a factor of 4 smaller than the linewidth in our high
mass core. \cite{wu.et.al05} showed that JCMT\,18354S exhibits
distinct blue-red asymmetries in the optically thick HCO+ and HCN
lines, both known to be a good tracers of infalling motions in
low-mass star-formation regions. A semi-analytic treatment of the
radiative transfer led them to conclude that the densities,
velocities and velocity dispersions implied by the observations
are, as to be expected, much larger than for low mass sources.

\begin{figure*}
    \centering
    \includegraphics[width=14 cm]{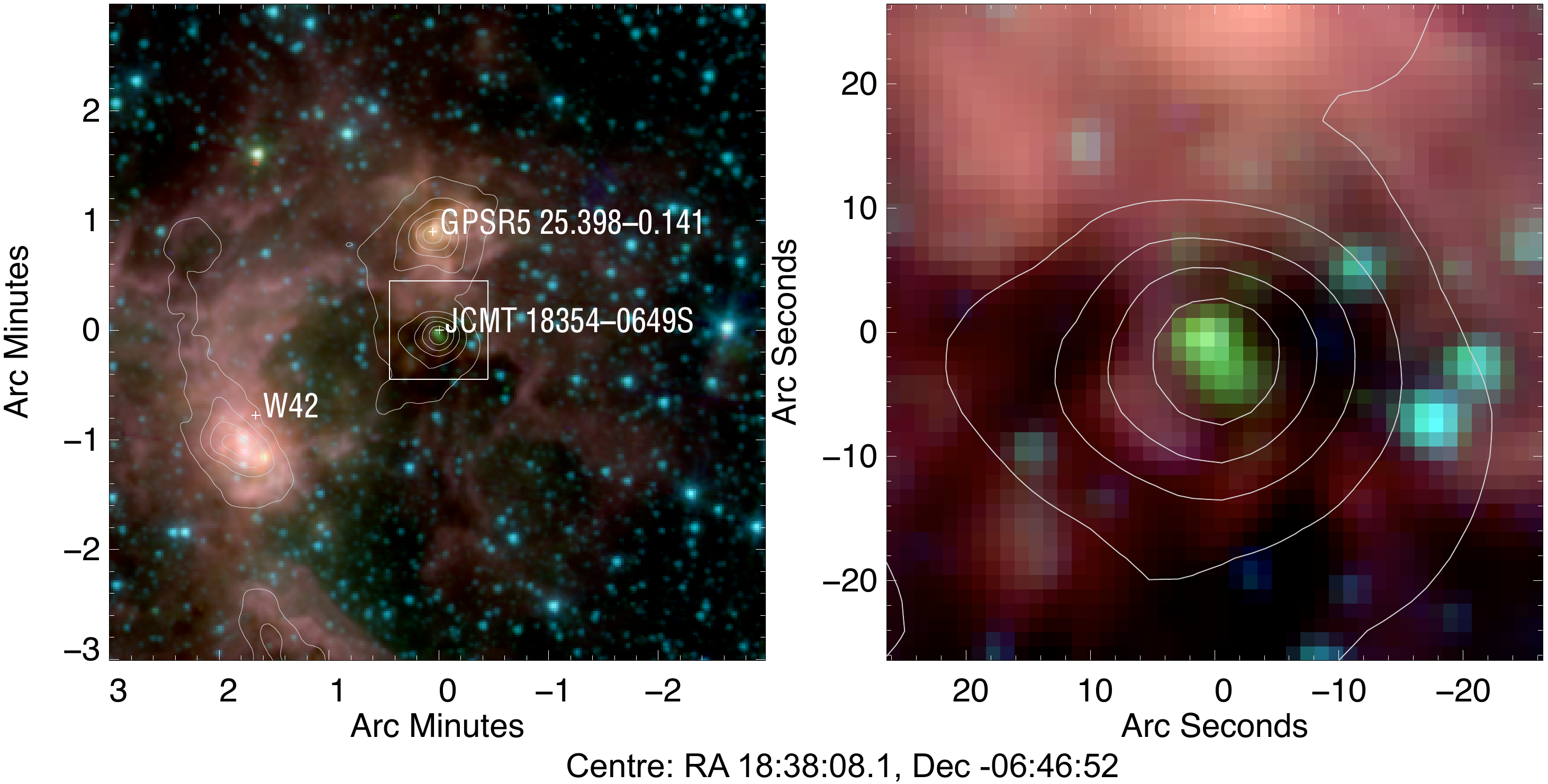}
    \caption{A colour composite image of IRAC photometric bands where Red is IRAC 8.0$\mu$m, Green is 4.5$\mu$m and the Blue is 3.6$\mu$m. Images are logarithmically scaled and combined into one composite image. The area enclosed in a white rectangle is the target source JCMT I18354-0649S and the image on the right is a close up of this region. GPSR5 25.398-0.141 and W42 are nearby UCH\textsc{ii} regions. In both images contours are from 850$\mu$m continuum JCMT SCUBA data. The contours start at 3$\sigma$ and increase by 1$\sigma$ thereafter. The position of each object and the scale are indicated on the image.}
    \label{Fig:rgb-spitzer}
\end{figure*}

Previous observations by \cite{wu.et.al05} determined that
JCMT\,18354S core has a size of 10" corresponding to 0.28pc at a
distance of 5700 pc away with a v$_{\rm lsr}$ of 95 km s$^{-1}$.
The total mass of 910 M$_{\odot}$ and 820 M$_{\odot}$ when derived
from 850 $\mu$m and 450 $\mu$m emission. The dust temperature was
estimated as \emph{T}$_{\rm d} = 14.4 $K by fitting a grey body to
the observed SED and assuming a dust emissivity index of $\beta =
2$ \citep{wu.et.al05}. However the dust temperature ranges from 11
to 29 K when uncertainties associated with the SCUBA fluxes and
dust emissivity index are taken into account. \cite{wu.et.al05}
fitted the line spectra of HCO$^{+} (3-2)$ and HCN$^{+} (3-2)$
with an analytic model from \cite{myers96}. The best fit model
gave a gas kinetic temperature \emph{T}$_{\rm K}$ of 16.7 K and
17.4 K for HCO$^{+} (3-2)$ and HCN$^{+} (3-2)$ respectively. This
agrees with a gas kinetic temperature of 17.5 K derived from
NH$_{3}$ emission by \cite{wu.et.al06}.

In this work we aim to characterise as fully as possible the
properties of JCMT\,18354S. To carry this out we present new
molecular line and archive \emph{Spitzer} observations of
JCMT\,18354S in Section~\ref{observations}. In
Section~\ref{radiative-transfer} we describe the 3D radiative
transfer code that is used to investigate these observations.
Using a radiative transfer code to model grids of line profiles
from different species and transitions means it is possible to
fully characterise both the global dynamical processes and the
local physical conditions in the gas in JCMT\,18354S. This then
allows for a comparison of the properties of JCMT\,18354S, a
massive core, with those of low mass cores
(Section~\ref{discussion}). In Section~\ref{conclusions} it is
concluded that many of the physical and chemical properties of
JCMT\,18354S are remarkably similar to those of a typical low mass
core. The key difference is in the degree of turbulence, with this
cold massive core exhibiting much more highly supersonic line
widths than in low mass cores, as would be expected from the
virial theorem.

\section{Observations}
\label{observations}

For our study of JCMT\,18354S, observations of both molecular line
and dust continuum emission are examined. The optically thin
continuum emission is used to infer physical parameters such as
the density of molecular gas. In addition the chemical and
dynamical processes occurring in JCMT\,18354S are investigated
through molecular spectral line data from different transitions.
The molecular spectral line data for JCMT\,18354S were obtained
using the JCMT and Mopra Telescope during several runs as listed
in the Table~\ref{tab:moldata}.

IRAC \citep{fazio-et-al-04} images of the region in all four
photometric bands were acquired from archived \emph{Spitzer} data.
The basic flux calibrated imaging data of the \emph{Spitzer}
Science Center (SSC) pipeline were used for further data reduction
and analysis. Cosmetic corrections and astrometric refinement were
performed with the MOPEX software \citep{markovoz-marleau-04} and
the final images (Fig.~\ref{Fig:rgb-spitzer}) were mosaicked using
scripts in the STARLINK reduction package.

Sub-millimetre continuum SCUBA pipeline reduced and calibrated
mosaics at 450$\mu$m and 850$\mu$m were acquired from the JCMT
archive\footnote{The JCMT Archive project is a collaboration
between the Canadian Astronomy Data Centre, Victoria and the James
Clerk Maxwell Telescope, Hilo}. The observations were taken in
September 2003 using a 64 point jiggle pattern to ensure
sufficient coverage of the sky. The sky opacity was estimated from
skydip observations giving an average $\tau_{850}$ of 0.3 and
$\tau_{450}$ of 1.6 and observations of Uranus taken on the same
night in photometry mode were used to derive Flux Calibration
Factors (FCF) of 244 Jy V$^{-1}$ and 344 Jy V$^{-1}$ for the
850$\mu$m and 450$\mu$m maps respectively.

Figure~\ref{Fig:rgb-spitzer} shows a colour composite view of the
region constructed from \emph{Spitzer} IRAC images. The overlaid
contours from 850$\mu$m JCMT SCUBA data highlight the structure
and the extent of the dust emission in the area. JCMT\,18354S is
situated in close proximity to the prominent UCH\textsc{ii} region
GPSR5\,25.398-0.141 to the north and W42 to the south-east.
\cite{wu.et.al05} report that JCMT\,18354S is only detectable in
sub-millimetre and millimetre wavelengths and suggest this is
indicative of its early stage of evolution. However close
examination of Fig.~\ref{Fig:rgb-spitzer} reveals an object
visible in all IRAC broad bands  at the very same position as the
sub-millimetre core (this data will have been unavailable to
\cite{wu.et.al05}. The new feature detected only in \emph{Spitzer}
IRAC bands is strongest in 4.5$\mu$m and could well indicate
emission originating from numerous H$_2$ lines present in those
wavelengths \citep{smith05,ybarra09} which would make JCMT\,18354S
a so-called 'green fuzzy'  object \citep{chambers09}. IRS spectra
will be required to confirm the nature of the shocked feature.

\begin{table*}
    \centering
    \caption{A list of the lines and frequencies of each transition that were observed. Different molecular gas species and transitions are observed in order to probe the varying physical conditions expected in protostellar clouds.}
    \label{tab:moldata}
    \begin{tabular}{lllcccc}
        \hline
        Line&
        Telescope&
        Date&
        Detector&
        $\nu_0$ (GHz)&
        HPBW $\left(^{\prime\prime}\right)$ &
        Map Type\\
        \hline
HCN $(J = 1~$--$~0)$&Mopra&05 Jul 2008&MOPS&88.63&36 & Raster Map\\
HCN $(J = 3~$--$~2)$&JCMT &20 Apr 2008&RxA &265.88&18.3 & 5 $\times$ 5 Map\\
HCO$^{+}$ $(J = 1~$--$~0)$&Mopra&05 Jul 2008&MOPS&89.19&36&Raster Map\\
HCO$^{+}$ $(J = 3~$--$~2)$&JCMT&27 Aug 2007&RxA&267.55&18.3&9 cross\\
H$^{13}$CO$^+$ $(J = 1~$--$ 0)$&Mopra&05 Jul 2008&MOPS&86.75&36&Raster Map\\
H$^{13}$CO$^+$ $(J = 3~$--$ 2)$&JCMT&09 May 2004&RxA&260.25&18.3&4 cross\\
$^{12}$CO $(J = 3~$--$~2)$&JCMT&05 Apr 2005&RxB&345.79&14.0& Raster Map\\
$^{13}$CO $(J = 1~$--$~0)$&Mopra&04 Jul 2008&MOPS&110.20&33&Raster Map\\
C$^{17}$O $(J = 3~$--$~2)$&JCMT&08 Oct 2004&RxB&337.06&14.6&3 $\times$ 3 Map\\
C$^{17}$O $(J = 2~$--$~1)$&JCMT&27 Aug 2007&RxA&224.71&21.5&5 strip\\
C$^{17}$O $(J = 1~$--$~0)$&Mopra&04 Jul 2008&MOPS&112.36&33&Raster Map\\
C$^{18}$O $(J = 1~$--$~0)$&Mopra&04 Jul 2008&MOPS&109.78&33&Raster Map\\
C$^{18}$O $(J = 2~$--$~1)$&JCMT&10 May 2004&RxA&219.56&21.3&5 strip\\
SiO $(J = 6~$--$~5)$&JCMT&25 Jun 2004&RxA&260.52&13.4&5 cross\\
SiO $(J = 2~$--$~1)$&Mopra&04 Jul 2008&MOPS&86.85&36&Raster Map\\
        \hline
    \end{tabular}
\end{table*}

Molecular line observations of JCMT\,18534S were obtained in
August of 2007 and April of 2008 with the JCMT using RxA receiver
operating from 211 to 279GHz frequencies and ACSIS autocorrelator
system in a position switching mode. The observations are spaced
at 10" to Nyquist sample the $\sim$ 20" full with of the JCMT beam
at these frequencies, an the main beam efficiency was 0.69. We
followed the the strategy described in \citep{wu.et.al05} during
their 2004 observing campaign (listed in the
Tab.\ref{tab:moldata}) choosing the same reference OFF-position
(800", 800") since it was free of emission at the frequencies
observed (HCN, HCO+, H$^{13}$CO). Telescope pointing was regularly
checked using the UCH\textsc{ii} region GPSR5\,25.398-0.14
situated to the North.

Our observations were supplemented with JCMT archive data of the
JCMT\,18534S taken with the RxA and RxB (325 to 375GHz) receivers
and DAS autocorrelator. In all cases the same type of
position-switching mode was implied apart from the $^{12}$CO $(J =
3~$--$~2)$ line where the raster-mapping mode was utilised. The
typical system temperatures (T$_{sys}$) for RxA and RxB receivers
are about 200K and 600K respectively. All the archival data was
reduced using the {\sc SPECX} package incorporated in the STARLINK
suite. Data taken with the new ACSIS autocorrelator were processed
using tasks in packages {\sc SMURF} and {\sc KAPPA} from the
STARLINK suite.

Additional spectral line data were obtained in July 2008 with the
Mopra Telescope, operated by the Australia Telescope National
Facility (ATNF). The observations were taken with the Mopra
Spectrometer (MOPS) in zoom mode to obtain on-the-fly (OTF) maps.
Each $3 \times 3$ arc-min OTF map took 30 minutes to cover and
pointing checks on SiO maser sources were carried out every hour.
Several OTF maps were made at each tuning to improve the signal to
noise by increasing the integration time. To reduce the effect of
scanning artefacts the telescope alternated between scanning in
right ascension and declination. The data were reduced with the
Livedata and Gridzilla packages \citep{barnes-et-al01}, which
corrects the bandpass for the off source reference spectra and
fits a polynomial baseline to the data. Miriad was used to smooth
the spectra from the original 0.1 km s$^{-1}$ pixels, by a 7-point
Hanning function, to improve the signal-to-noise, and resampled to
0.2 km s$^{-1}$ pixels (Nyquist sampled). The beam-size, T$_{sys}$
and efficiency of the Mopra telescope is 36 arc-sec, 200K and 0.49
at 86 GHz and 33 arc-sec, 600K and 0.42 at 115 GHz
\citep{ladd-et-al-05}. All spectra presented in
Figures~\ref{Fig:c17o-2-1}~--~\ref{Fig:model-12co} indicate their
position as an offset from the same common centre which is centred
on the 850 $\mu$m emission peak of JCMT\,18534S at RA =
18$^{\rm{h}}$ 38$^{\rm{m}}$ 8.1$^{\rm{s}}$, Dec = -6$^{\circ}$
46$^{\prime}$ 52$^{\prime\prime}$. In all cases the mapping offset
positions are indicated on the individual figures.

\section{Modelling}
\label{sect:modeling}

In the following sections, a model for JCMT\,18534S is carefully
developed. Firstly a semi-analytic depletion analysis is presented
in which the degree of freeze-out of molecules onto dust grains is
estimated. This is shown to be consistent with the full radiative
transfer model which follows.  Secondly, the dynamical and
radiative transfer model is presented. Many processes (freeze-out,
desorption, infall, outflow) occur simultaneously in cold dark
molecular clouds. Observed line profiles are made up of emission
from gas that is susceptible to many or all of these effects.
Different molecular transitions preferentially trace different
physical regimes and dynamical phenomena. Including all processes
in a consistent model is a difficulty that can be overcome by
using a 3D molecular line radiative transfer code. Model fits to
observational line profiles are presented in a sequence that
allows the dynamical elements of the model to be explained in
turn.

The 3-D radiative transfer code used throughout this work was
written and developed by Keto and collaborators \citep[see,
e.g.][for examples of its
use]{keto90,keto.et.al04,redman.et.al04a,redman06,carolan-et-al-08}.
The code will shortly be made public under the name {\sc mollie}
(MOLecular LIne Explorer) and is used to generate synthetic line
profiles to compare with observed rotational transition lines. In
order to calculate the level populations the statistical
equilibrium equations are solved using an Accelerated Lambda
Iteration (ALI) algorithm \citep{rybicki} that reduces the
radiative transfer equations to a series of linear problems that
are solved quickly even in optically thick conditions. {\sc
mollie} splits the overall structure of a cloud into a 3-D grid of
distinct cells where density, abundance, temperature, velocity and
turbulent velocity are defined.

The best fit parameters are shown in Tab\,\ref{lines} where the
number density are in bold font to indicate that it is the peak
value of a profile that is shown in Fig.~\ref{Fig:param-profiles}.
It is created from the observations of SCUBA 850$\mu$m emission
with the assumption of spherical symmetry. Initially the intensity
drops off as $r^{-1.5}$ progressing to a steeper value of
$r^{-2}$. This indicates the presence of a core which is
surrounded by a gradually diffuse outer region. The beam size at
850$\mu$m is $14.5^{\prime\prime}$ and the intensity used to
derive the density is averaged over the size of the beam.

The velocities in Tab\,\ref{lines} are of gas that is moving
towards the centre of the core (Infall) except for the velocity of
$^{12}$CO (Outflow) whose direction is away from the core centre.
The turbulent motion of the gas is characterised with the
turbulent velocity parameter. The abundance is the ratio of the
column density with respect to the column density of H$_{2}$ or
$N_{\rm species}/N_{\rm H_{2}}$.

\subsection{Freeze-out of Molecular Gas onto Dust Grains}
\label{sect:optically-thin}

While molecular line rotational transitions are a very powerful
probe of cloud conditions, continuum observations can also be used
to unambiguously investigate some processes. The continuum
radiation from dust grains is optically thin in the wavelengths
observed in this paper and are used in tandem with molecular line
observations to investigate the freeze-out of molecular gas onto
dust grains. Freeze-out of molecular gas onto dust grains is
expected to occur when the temperature of protostellar clouds is
low $T < 20$K and density is high $n_{\rm{H}_{2}} > 10^{4}$
cm$^{-3}$ \citep{sandford}. During this phase it is difficult to
observe emission from rotational transitions. However molecular
hydrogen is the dominant constituent of molecular clouds and
distinct relationships exist between it and the next most dominant
molecule in molecular clouds which is carbon monoxide and its
isotopes. With this knowledge the column density of H$_{2}$ is
inferred from dust and gas emission and a discrepancy between the
two indicates that there is significant freeze-out of molecular
gas.

\subsubsection{Column Density from Dust Emission}

The dust emission at 850~$\mu$m is optically thin therefore it is
used to infer the H$_{2}$ column density in cm$^{-2}$ across the
cloud using the equation

\begin{equation}
N_{\textrm{H}_{2}} = \frac{S_{850}}{\Omega \kappa_{850} \mu
m_{\rm{H}} B(\textrm{T})}. \label{column-den-from-dust}
\end{equation}

$S_{\rm{850}}$ is the flux in Jy from dust emission maps at
850$\mu$m. The flux is measured in an aperture of
21$^{\prime\prime}$ which is the size as the main beam of
C$^{17}$O and C$^{18}$O $(2~$--$~1)$ observations. $\Omega$ is the
aperture solid angle in steradians; $\Omega = (\pi \Theta^{2})/(4
\ln 2)$, where $\Theta$ is the beam size in radians. The Planck
function $B(\textrm{T})$ also in Jy is calculated assuming a dust
temperature of 20K, $m_{\rm{H}}$ is the mass of a hydrogen atom in
grams and $\mu$ is the mean molecular weight. A dust mass opacity
of $\kappa_{850}$ = 0.02 cm$^{2}$~g$^{-1}$ is assumed which is
consistent with \cite{ossenkopf&henning94} and it represents dust
grains with thin ice mantles, the dust to gas ratio is taken to be
100.

\subsubsection{Column Density from Line Emission}

The column density of H$_{2}$ was calculated using optically thin
gas emission line observations of the transitions C$^{18}$O $(J =
2~$--$~1)$, C$^{17}$O $(J = 2~$--$~1)$ and C$^{17}$O $(J =
3~$--$~2)$. This is done to compare the emission from both
continuum and line species to observe the degree, if any, of
molecular gas freeze-out. For an optically thin gas with
negligible contribution from the cosmic microwave background, the
column density of a species is given by
\begin{equation}
N_{\rm spec} = 1.67 \times
10^{14}~\textrm{cm}^{-2}~\frac{Q(\textrm{T}_{\textrm{ex}})}{\nu
\mu^{2} S} \times \exp \left(\frac{E_{\textrm{u}}}{k
T_{\textrm{ex}}} \right) \int T_{\textrm{mb}}~\textrm{d}V
    \label{column-den-from-gas}
\end{equation}
where $\nu$ is the frequency of the transition in GHz, k is
Boltzmann's constant, $\int T_{\rm{mb}}~\rm{d}V$ is the integrated
intensity of the line profile, $S$ is the line strength,
E$_{\rm{u}}$ is the energy of the upper level and $Q(T_{\rm{ex}})$
is the partition function. An excitation temperature of 20K is
assumed which is consistent with the dust temperature and for
C$^{18}$O and C$^{17}$O. The partition function is 7.94 and 7.69
respectively. $\mu$ is the dipole moment in Debyes which for
C$^{18}$O is 0.11079 Debye and for C$^{17}$O  it is 0.11034 Debye
\citep{pickett.et.al98}. The conversion factor of
C$^{18}$O~$\rightarrow$~H$_{2}$ is 2.07~$\times$~10$^{6}$ and
C$^{17}$O~$\rightarrow$~H$_{2}$ is 7.56~$\times$~10$^{6}$
\citep{ladd.et.al98,lacy.et.al94,wilson_rood_94}.

\subsubsection{Column Density from a Constant Abundance Model}
\label{radiative-transfer}

As a further constraint on the degree of freeze-out the column
density of H$_{2}$ is determined using {\sc mollie}. While {\sc
mollie} can implement any type of 3D physical structure, a
spherical model is used to approximate the cloud of gas in
JCMT\,18354S for this calculation only. A constant abundance model
is run and from this we have determined the predicted column
density of H$_{2}$ to compare with the column density of H$_{2}$
calculated from the dust emission. The figures showing the
constant abundance model fit to the observed line profiles are not
shown in this paper, we only wish to highlight the error of using
a constant abundance model.

A best fit model was found which matched the peak intensity and
width of the line profiles of C$^{18}$O (2 - 1), C$^{17}$O (2 - 1)
and C$^{17}$O (3 - 2). The density of H$_{2}$ has a peak value of
10$^{6}$~cm$^{-3}$ and varies as $r^{-2}$ from the centre of the
model cloud. This density profile is similar as that observed from
low mass Class 0 protostars by \cite{shirley}. The radius of the
core was set at 30$^{\prime\prime}$ and is derived from 850 $\mu$m
emission which is at the 3$\sigma$ level at this point. This
radius is the same as where molecular line emission from high
critical density species such as HCN (3 - 2) and HCO$^{+}$ (3 - 2)
is just indistinguishable from the continuum. However there is
significant emission observed beyond this radius from gas species
with low critical densities.

It was found that an abundance of $1.4 \times 10^{-8}$ and $1.0
\times 10^{-9}$ for C$^{18}$O and C$^{17}$O gave the best fit to
the observed line profile. However these values are an order of
magnitude less than estimated using the standard interstellar
abundance ratios of H$_{2}$/CO = 3700, $^{16}$O/$^{18}$O = 560 and
$^{18}$O/$^{17}$O = 3.65
\citep{ladd.et.al98,lacy.et.al94,wilson_rood_94}. This result
agrees with the outcome of the previous section where the observed
integrated intensity gave a H$_{2}$ column density that is
significantly less than that derived from optically thin dust
emission.

\subsubsection{Comparing the Dust and Gas Emission}
\label{gasdast-comparison}

\begin{figure}
    \centering
    \includegraphics[width=8cm]{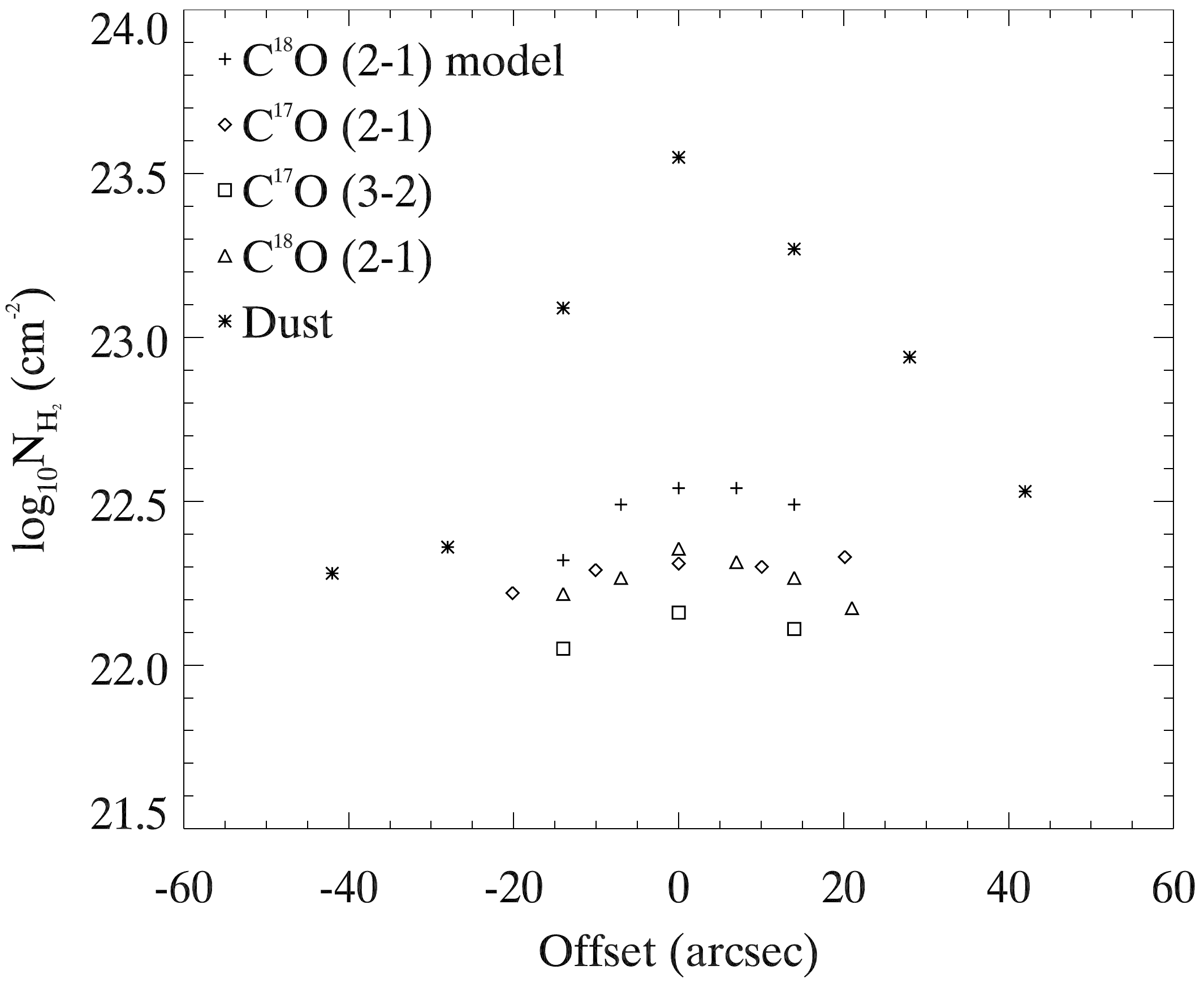}
    \caption{The column density of H$_{2}$ was calculated with observations of dust continuum and molecular line emission, in addition the column density of H$_{2}$ was calculated from the best fit model to the line profiles of C$^{18}$O $(J = 2~$--$~1)$ assuming a constant abundance model. It indicates that significant amounts of the molecular gas is not observed because the gas has frozen-out onto dust grains. Therefore an abundance profile is included in the modelling to account for this effect.}
    \label{Fig:freezeout-i18354}
\end{figure}

The column density of H$_{2}$ using all three techniques is plotted in Fig.~\ref{Fig:freezeout-i18354} as a function of distance from the centre of the core. There is an order of magnitude difference between the column density of H$_{2}$ calculated from dust emission, several optically thin molecular gas species, and the model. This is caused by freeze-out of molecular gas onto dust grains. 
To incorporate this effect in {\sc mollie} a freeze-out radius is
defined. The extent of the freeze-out radius is estimated by
comparing the point where the H$_{2}$ column density derived from
dust and gas observations diverge. Referring to
Fig.~\ref{Fig:freezeout-i18354} this occurs at $\approx
20^{\prime\prime}$ (0.5 pc) from the centre of the core. Due to
the large beam of the observations this region is not sampled to a
high resolution which creates a sharp drop off in the abundance,
whereas an abundance determined from interferometric observations
would show a gradual drop off.

In choosing an abundance and the degree of freeze-out, we are
limited by the requirement that the H$_{2}$ column density derived
from gas observations should agree with the H$_{2}$ column density
derived from dust observations. In addition the modelled and
observed spectral line profile should agree. Therefore the
abundance used in our model is the abundance that gives a H$_{2}$
column density that agrees with the dust derived column density
but which is reduced by an order of magnitude inside the
freeze-out radius.

\begin{figure}
    \centering
    \includegraphics[width=7cm,height=7cm]{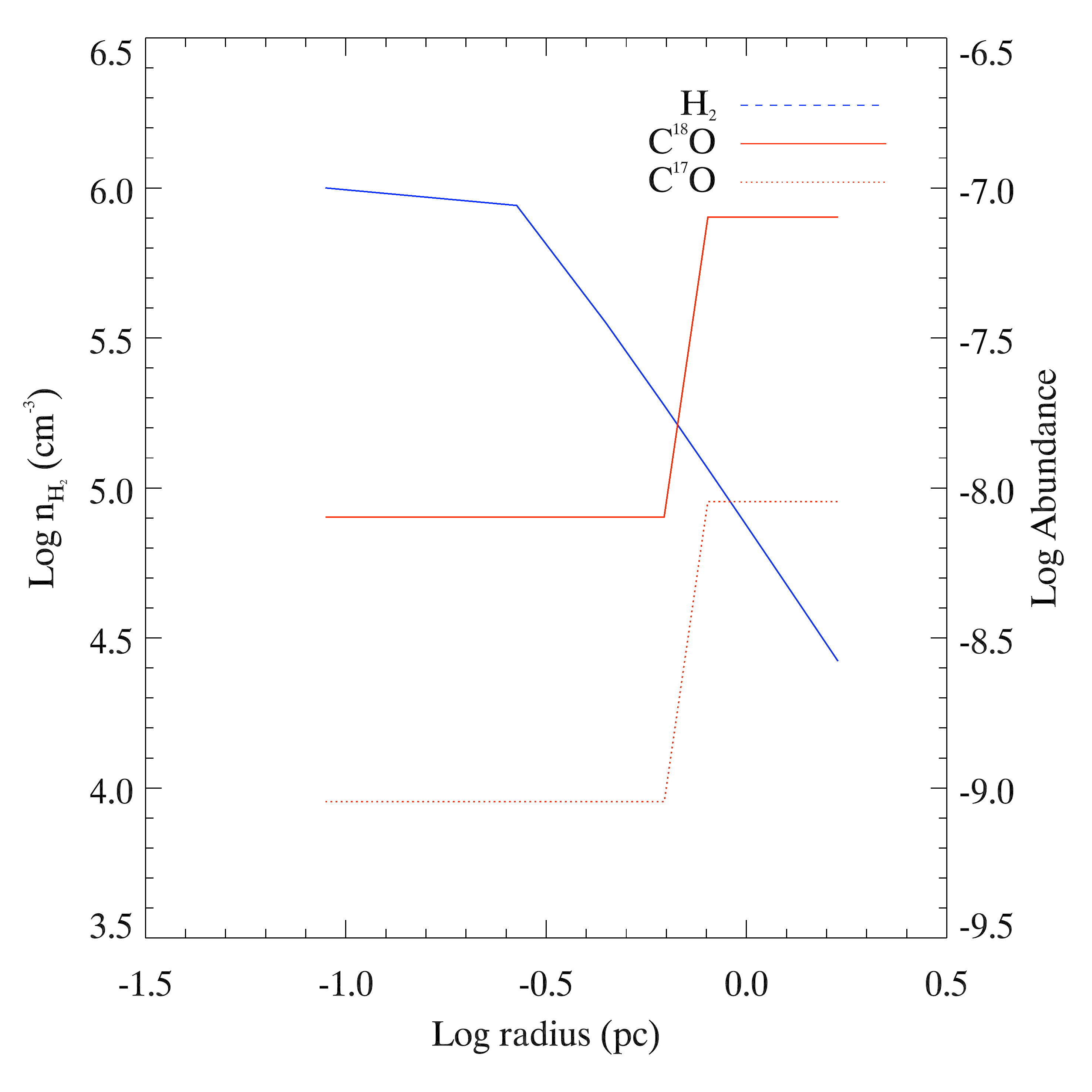}
    \caption{Profiles of the parameters that vary as a function of distance from the centre of the cloud. These profiles show how the H$_{2}$ density (blue) and molecular abundance (red) vary as a function of distance from centre of the cloud. The density of H$_{2}$ varies as a power law and the abundance decreases because of freeze-out.}
    \label{Fig:param-profiles}
\end{figure}

\begin{table*}
    \caption{Parameters for the best fit model. The abundance is the ratio of the column density with respect to the column density of H$_{2}$ or $N_{\rm species}/N_{\rm H_{2}}$. The two values of abundance quoted refer to their value inside and outside the freeze-out radius (Section \ref{gasdast-comparison}). Included in this table, for comparative purposes, is a list of the best fit parameters for the low mass protostellar core L483. The key feature of this comparison is that gas in cores forming high mass stars tends to have a higher infall velocity and a much higher turbulent velocity.}
    \begin{center}
        \begin{tabular}{@{}cccccl@{}}
        \hline
        Molecule&
        Density&
        Velocity&
        Temperature&
        Turbulent Velocity&
        Abundance\\
                &
        (cm$^{-3}$)&
        (km~s$^{-1}$)&
        (K)&
        (km~s$^{-1}$)&
        ($\times$ 10$^{-8}$)\\
\hline
JCMT 18354S (This work) & & & & & \\
C$^{17}$O           &\bf{10$^{6}$}&1&18&1.4& 0.09 and 0.9 \\
C$^{18}$O           &\bf{10$^{6}$}&1&20&1.6& 0.7 and 8\\
HCO$^{+}$           &\bf{10$^{6}$}&1.5&20&1.5& 0.022\\
H$^{13}$CO$^{+}$    &\bf{10$^{6}$}&1&20&1.5& 0.0002\\
HCN                 &\bf{10$^{6}$}&1.5&20&1.5& 0.01\\
$^{13}$CO           &\bf{10$^{6}$}&1.5&20&1.5& 90\\
$^{12}$CO (Envelope)&\bf{10$^{6}$}&1&20&1.6&   1000\\
$^{12}$CO (Outflow) &5 $\times$10$^{4}$&10&110&1.7& 6000\\
\hline
L483 \citep{carolan-et-al-08} & & & & & \\
C$^{17}$O           &\bf{10$^{6}$}&0.6&10&0.1&1\\
C$^{18}$O           &\bf{10$^{6}$}&0.6&10&0.1&5\\
$^{13}$CO           &\bf{10$^{6}$}&0.6&10&0.1&25\\
$^{12}$CO (Envelope)&\bf{10$^{6}$}&0.6&10&0.1& 1700\\
$^{12}$CO (Outflow) &$10^{4}$&3.5&90&1.4& 3500\\
        \hline
        \end{tabular}
    \end{center}
\label{lines}
\end{table*}

\subsection{Gas dynamics}

In this section, the inclusion of several global dynamical
processes in the model for JCMT\,18354S is justified. The
molecular line transition that best exhibits each process is shown
together with the model fit. It should be emphasised that the
model line profiles are from the final overall model  and that
some transitions are affected by several of the effects discussed
below. The model parameters are displayed in Table~\ref{lines} and
discussed in the Section~\ref{discussion}.

\subsubsection{Rotating Molecular Gas}

\begin{figure*}
    \centering
    \includegraphics[width=15cm]{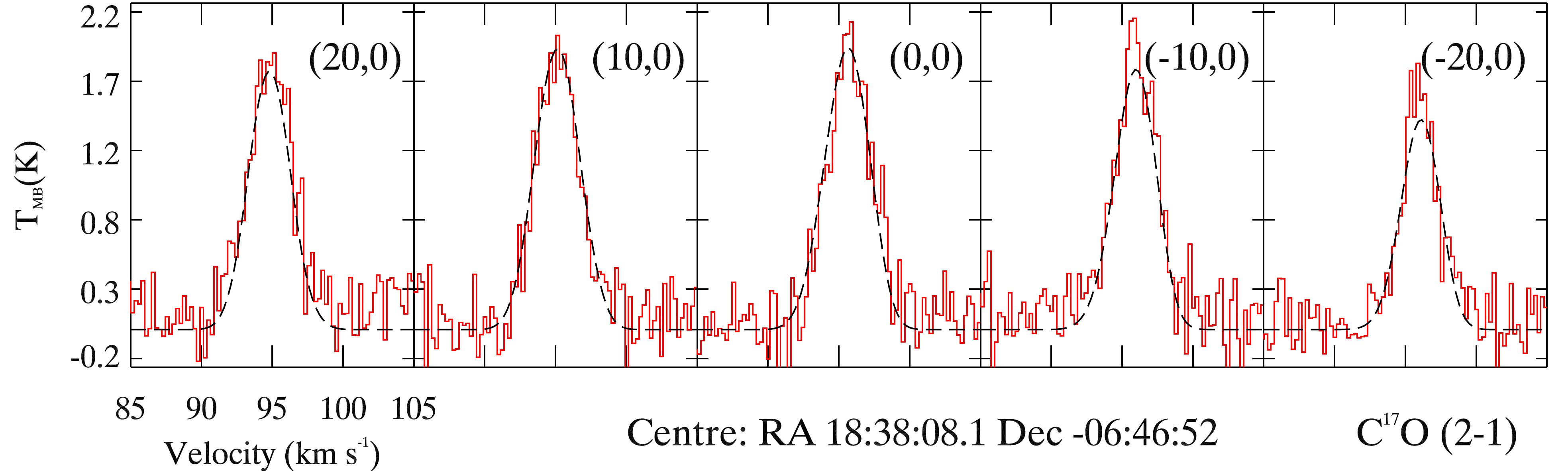}
    \caption{C$^{17}$O $(2~$--$~1)$ line spectra taken with the JCMT. The frequency of this transition is 224.71 GHz and the spectral resolution is 0.24~km~s$^{-1}$. The line profile positions are with respect to the centre of Fig.~\ref{Fig:rgb-spitzer} and each line profile is offset from each other by 10$^{\prime\prime}$. The red solid line is the data and the black dashed line is the model.}
    \label{Fig:c17o-2-1}
\end{figure*}

\begin{figure}
    \centering
    \includegraphics[width = 6 cm,height = 6 cm]{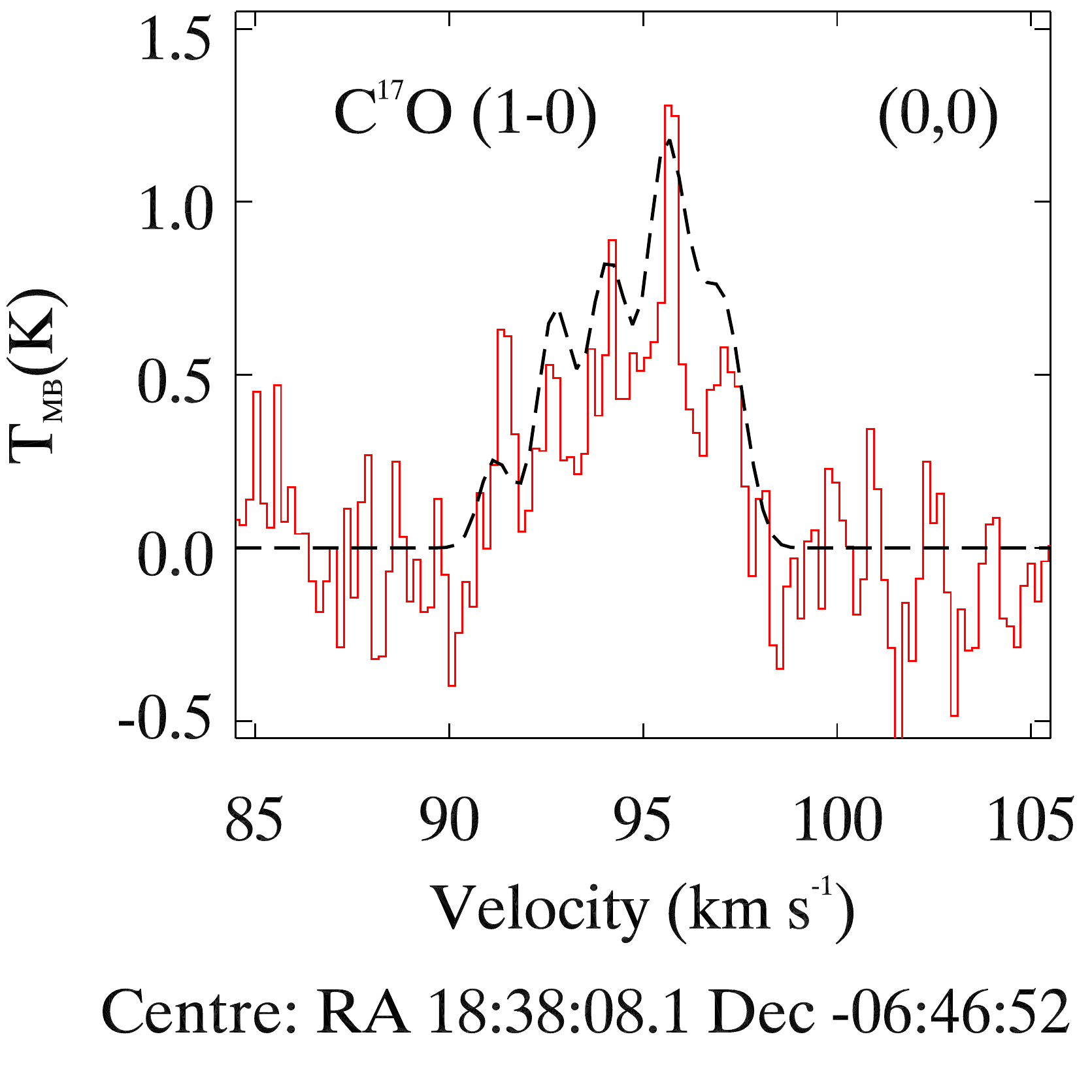}
    \caption{C$^{17}$O $(J = 1~$--$~0)$ line spectra taken with the Mopra telescope. The frequency of this transition is 112.36 GHz and the spectral resolution is 0.18~km~s$^{-1}$. The line profile positions are with respect to the centre of Fig.~\ref{Fig:rgb-spitzer}. The red solid line is the data and the black dashed line is the model.}
    \label{Fig:c17o-1-0}
\end{figure}

\begin{figure}
    \centering
    \includegraphics[width=8cm,height=8cm]{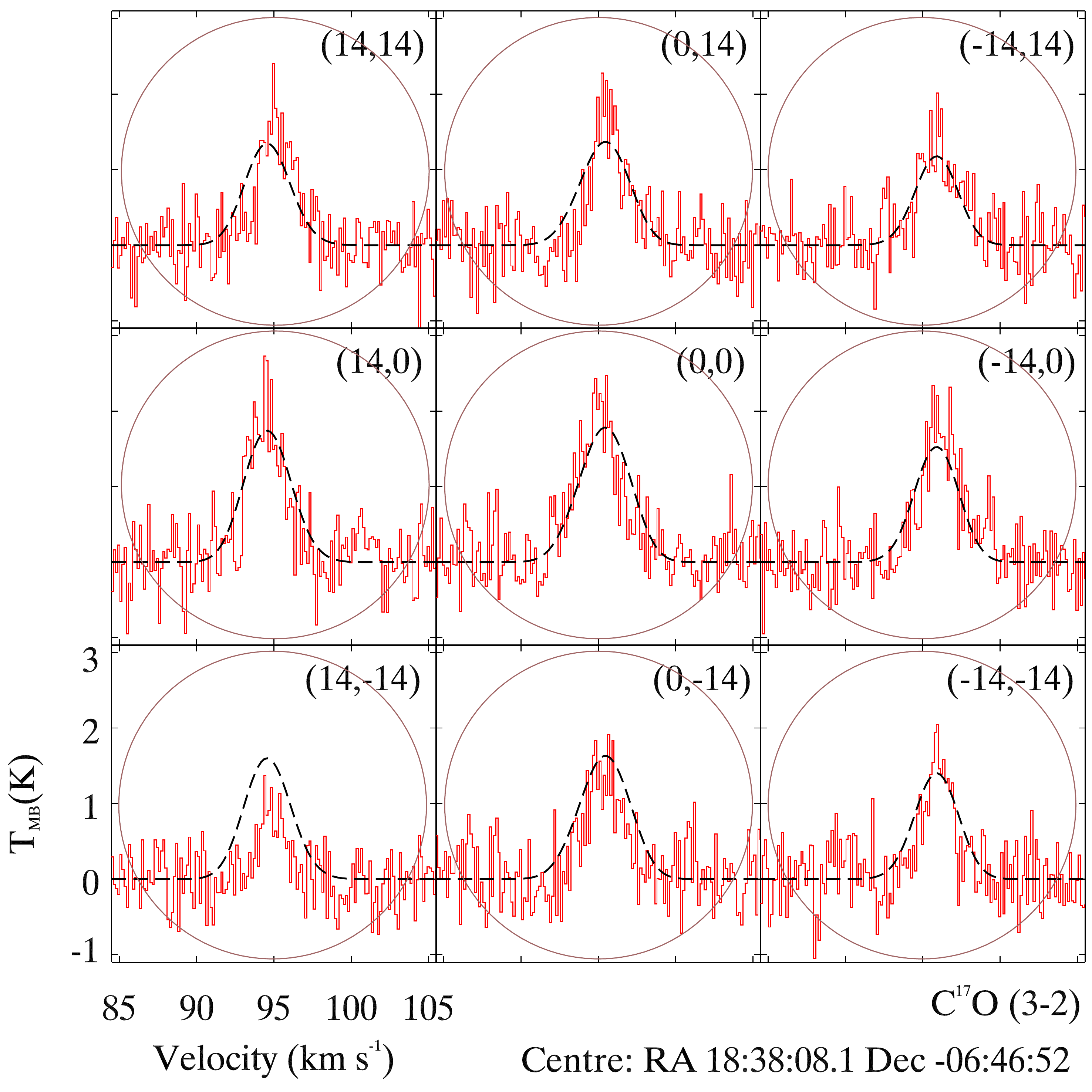}
    \caption{C$^{17}$O $(3~$--$~2)$ line spectra taken with the JCMT. The frequency of this transition is 337.06 GHz and the spectral resolution is 0.14~km~s$^{-1}$. The line profile positions correspond to the centre of Fig.~\ref{Fig:rgb-spitzer} and each line profile is offset from each other by 14$^{\prime\prime}$. The circles show the pointing pattern used in this transition only since it is under-sampled with respect to the JCMT beam.}
    \label{Fig:c17o-3-2}
\end{figure}

\begin{figure*}
    \centering
    \includegraphics[width=15cm]{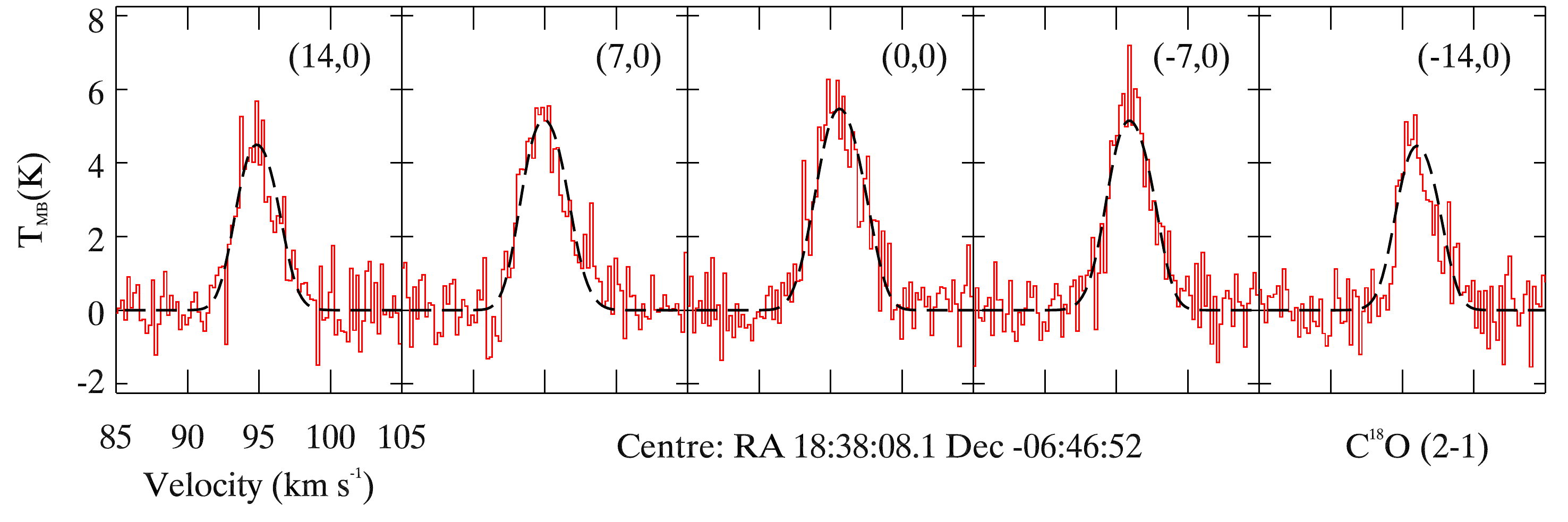}
    \caption{C$^{18}$O $(J = 2~$--$~1)$ line spectra taken with the JCMT. The frequency of this transition is 219.56 GHz and the spectral resolution is 0.24~km~s$^{-1}$. The line profile positions correspond to the centre of Fig.~\ref{Fig:rgb-spitzer} and each line profile is offset from each other by 7$^{\prime\prime}$.}
    \label{Fig:c18o-2-1}
\end{figure*}

\begin{figure}
    \centering
    \includegraphics[width = 6cm,height = 6cm]{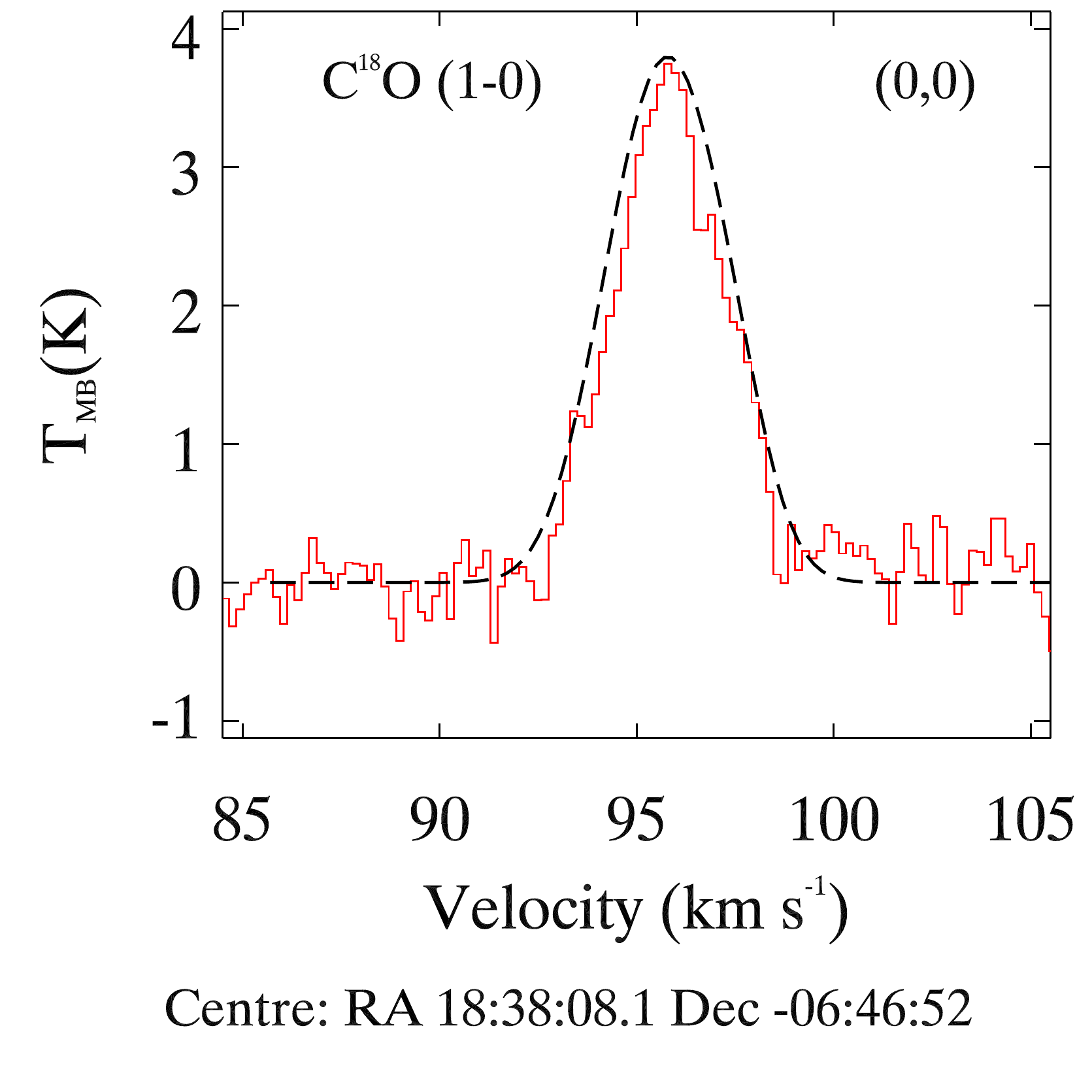}
    \caption{C$^{18}$O $(1~$--$~0)$ line spectra taken with the Mopra telescope. The frequency of this transition is 109.78 GHz and the spectral resolution is  0.18~km~s$^{-1}$. The line profile position corresponds to the centre of Fig.~\ref{Fig:rgb-spitzer}.}
    \label{Fig:c18o-1-0}
\end{figure}

The line emission spectra of three transitions in C$^{17}$O are
shown in Fig.~\ref{Fig:c17o-2-1}, \ref{Fig:c17o-1-0} and
\ref{Fig:c17o-3-2}. Alongside the observed line profiles (red) are
line emission profiles from the best fit model (black) created
using {\sc mollie}. The line profiles are created using a model
where the abundance varies due to freeze-out as discussed in the
previous subsection. Figure~\ref{Fig:param-profiles} shows the
abundance and density profile used to model the transitions of
C$^{17}$O and C$^{18}$O. The temperature and velocity of the
turbulent gas are kept constant throughout the cloud in this
model. The position of the peak intensity shifts in position on
the velocity axis. This shift is most readily seen in the
C$^{17}$O (Fig.~\ref{Fig:c17o-2-1}, \ref{Fig:c17o-1-0} and
\ref{Fig:c17o-3-2}) and C$^{18}$O (Fig.~\ref{Fig:c18o-2-1} and
\ref{Fig:c18o-1-0}) transitions. These lines show a shift in the
position of the peak intensity which is caused by rotating gas. It
is found that a rotational velocity of 0.5 km s$^{-1}$ in the
model gives the best fit to the observed line profile.


\subsubsection{Turbulent Molecular Gas}

Two rotational transitions of the C$^{18}$O molecule are shown in
Fig.~\ref{Fig:c18o-2-1} \&~\ref{Fig:c18o-1-0}. The line emission
profiles are single peaked and very broad ($\sim$ 3 km s$^{-1}$ )
with respect to line emission profiles of the same transition from
low mass protostellar cores \citep{carolan-et-al-08}. The best fit
model to the observed line profile includes gas with a high
turbulent velocity $\approx$ 1.5 km s$^{-1}$. This parameter is
constrained from a comparison of the modelled line emission
profile and the hyperfine line emission in the three transitions
of C$^{17}$O. The hyperfine lines are readily observed in low
turbulent velocity gas from low mass protostellar cores
\citep{redman.et.al02b}. Fig.~\ref{Fig:c17o-2-1},
\ref{Fig:c17o-1-0} and \ref{Fig:c17o-3-2} are actually composed of
hyperfine lines but instead they are blended together indicating
the gas is highly turbulent.

\subsubsection{Infalling Molecular Gas}

The detection of infalling molecular gas is one of the key
indicators of star forming activity in low mass protostellar
clouds. The envelope acts a reservoir for new material and the
radius within which gas is infalling grows with time. Optically
thick emission from gas creates its own distinctive line profile.
The "classic" signature of infalling gas is a blue asymmetric line
profile where the blue-shifted emission peaks at a higher
intensity than the red-shifted emission \citep{myers96}. The
asymmetry occurs because radiation from warm gas at the centre of
the cloud is absorbed by cooler gas that lies farther out. In
contrast the blue-shifted emission does not pass in front of an
absorbing layer of gas thereby creating an asymmetric line
profile.

\begin{figure}
    \centering
    \includegraphics[width=6cm]{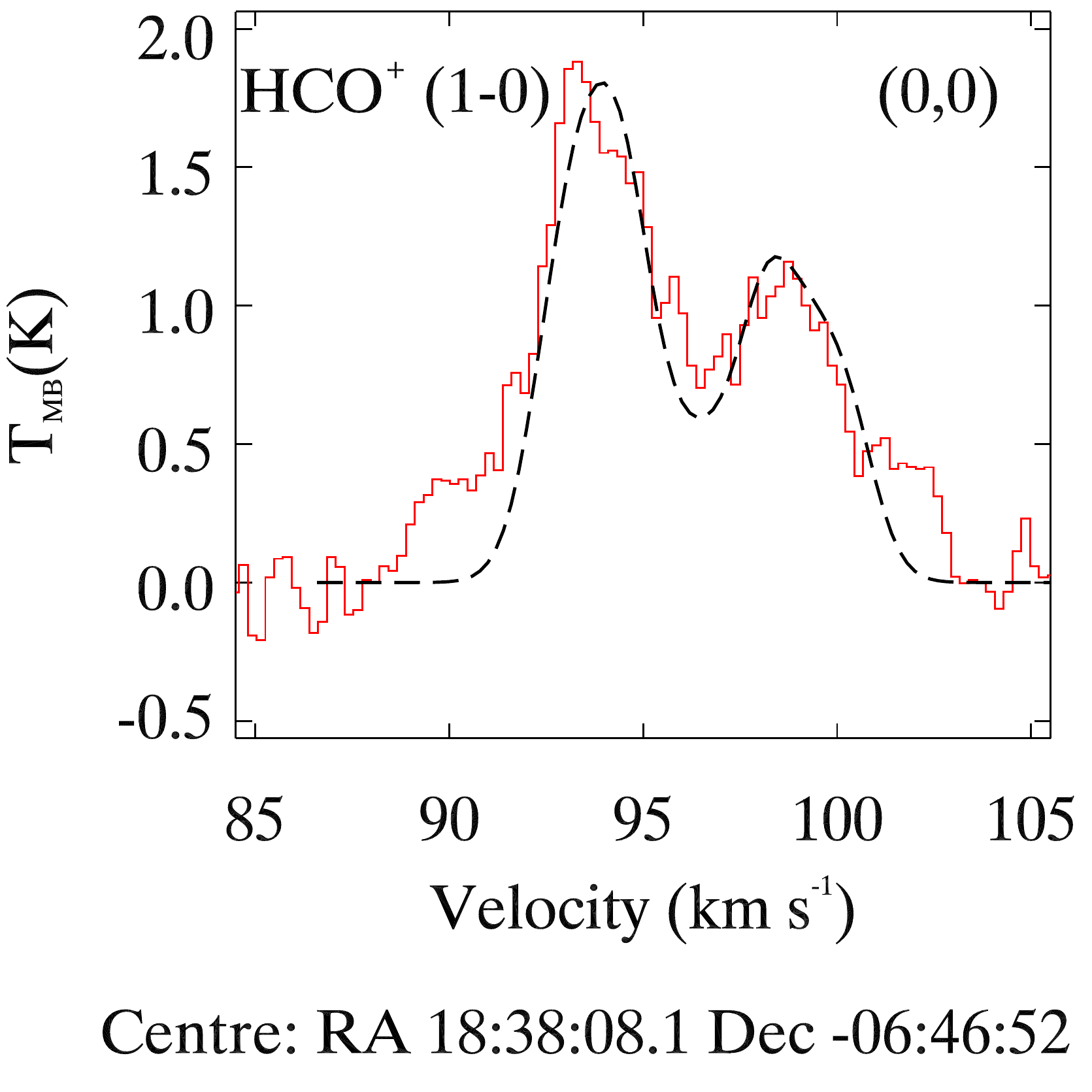}
    \caption{HCO$^{+}$ $(J = 1~$--$~0)$ line spectra taken with the Mopra telescope. The frequency of this transition is 89.19 GHz and the spectral resolution is 0.23~km~s$^{-1}$. The line profile position corresponds to the centre of Fig.~\ref{Fig:rgb-spitzer}.}
    \label{Fig:hcoplus-10}
\end{figure}

\begin{figure}
    \centering
    \includegraphics[width=10cm]{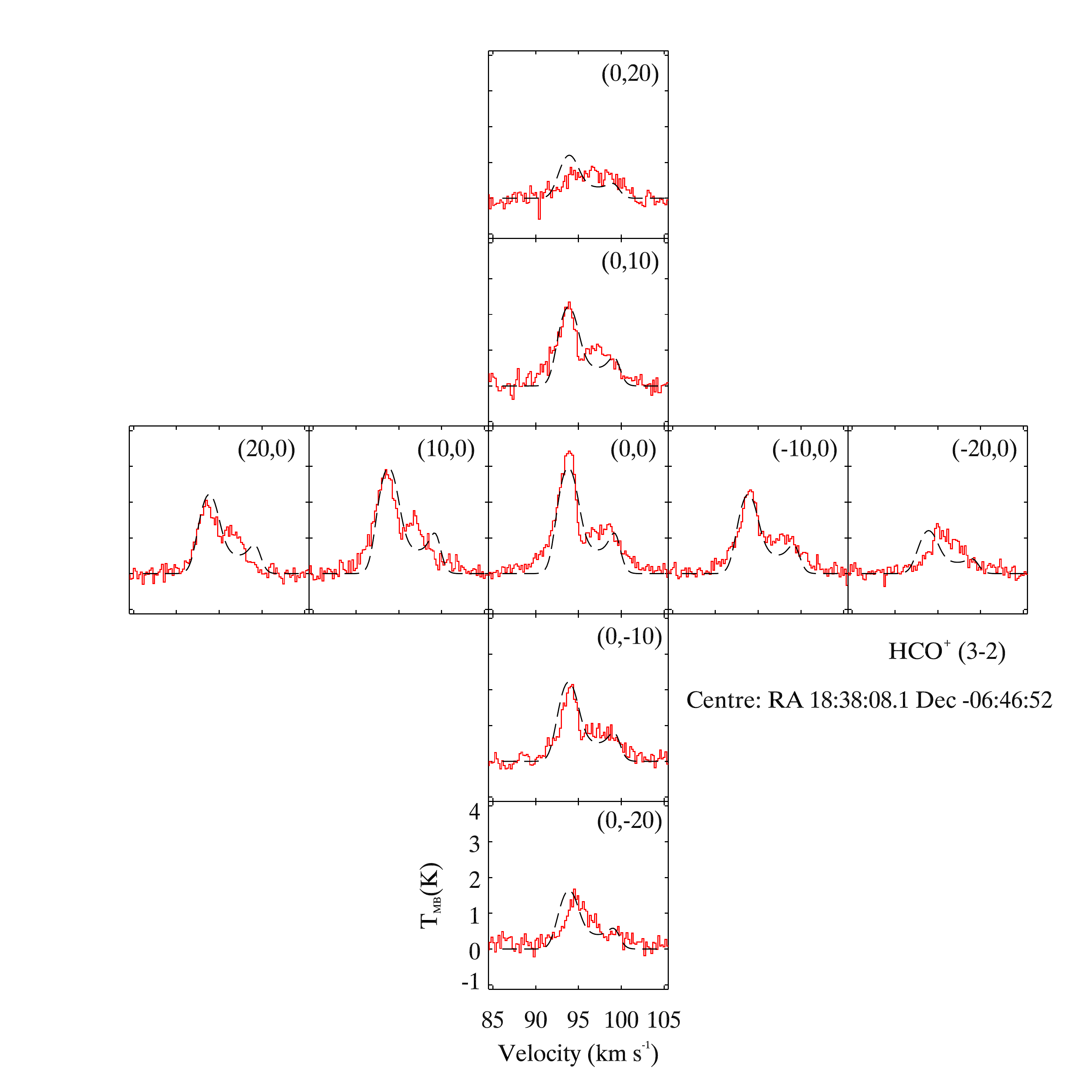}
    \caption{HCO$^{+}$ $(J = 3~$--$~2)$ line spectra taken with the JCMT. The frequency of this transition is 267.55 GHz and the spectral resolution is 0.09~km~s$^{-1}$. The line profile positions correspond to the centre of Fig.~\ref{Fig:rgb-spitzer} and each line profile is offset from each other by
10$^{\prime\prime}$.}
    \label{Fig:hcoplus-32}
\end{figure}

\begin{figure}
    \centering
    \includegraphics[width=6cm]{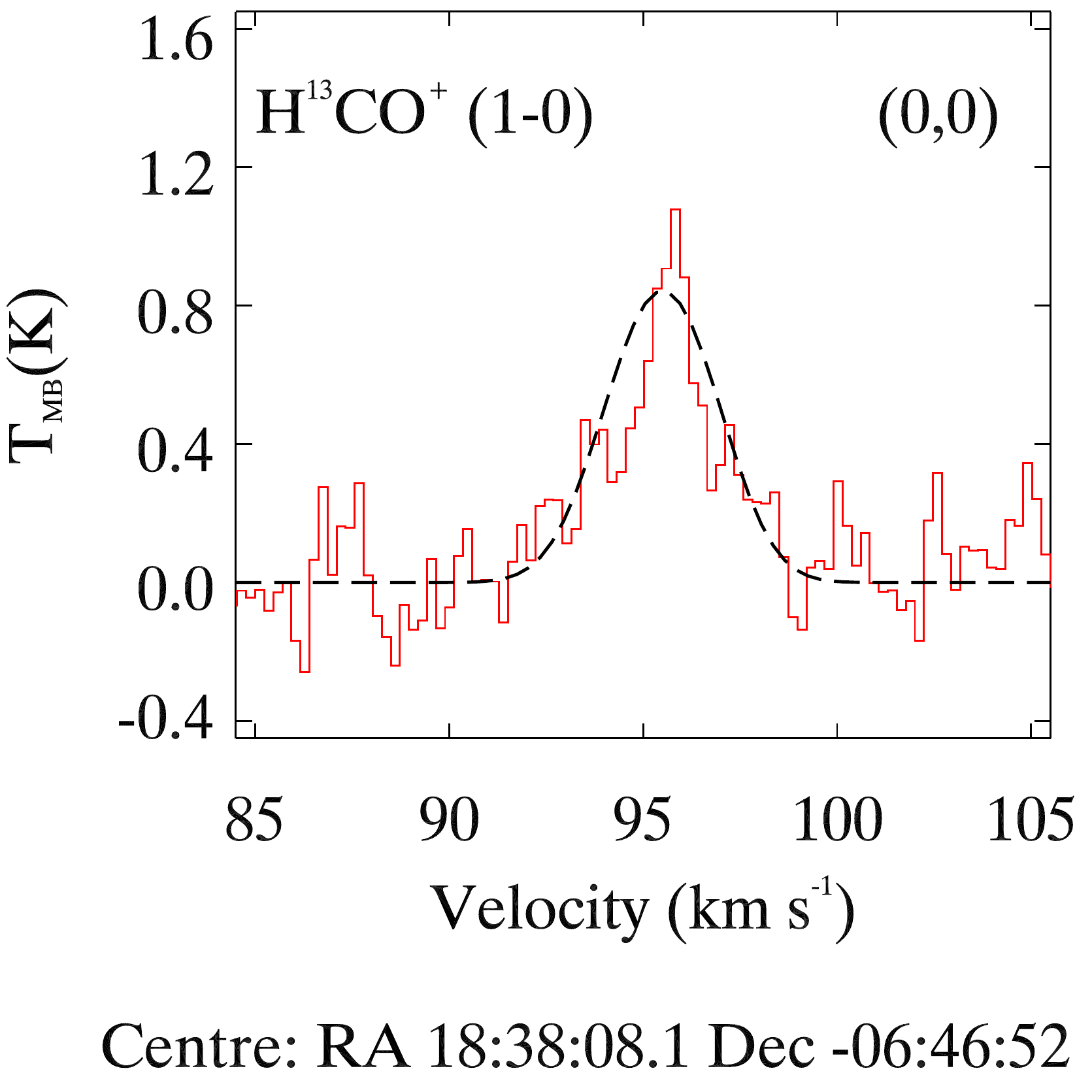}
    \caption{H$^{13}$CO$^{+}$ $(J = 1~$--$~0)$ line spectra taken with the Mopra telescope. The frequency of this transition is 86.75 GHz and the spectral resolution is 0.23~km~s$^{-1}$. The line profile position corresponds to the centre of Fig.~\ref{Fig:rgb-spitzer}.}
    \label{Fig:h13coplus-10}
\end{figure}

\begin{figure}
    \centering
    \includegraphics[width=8cm]{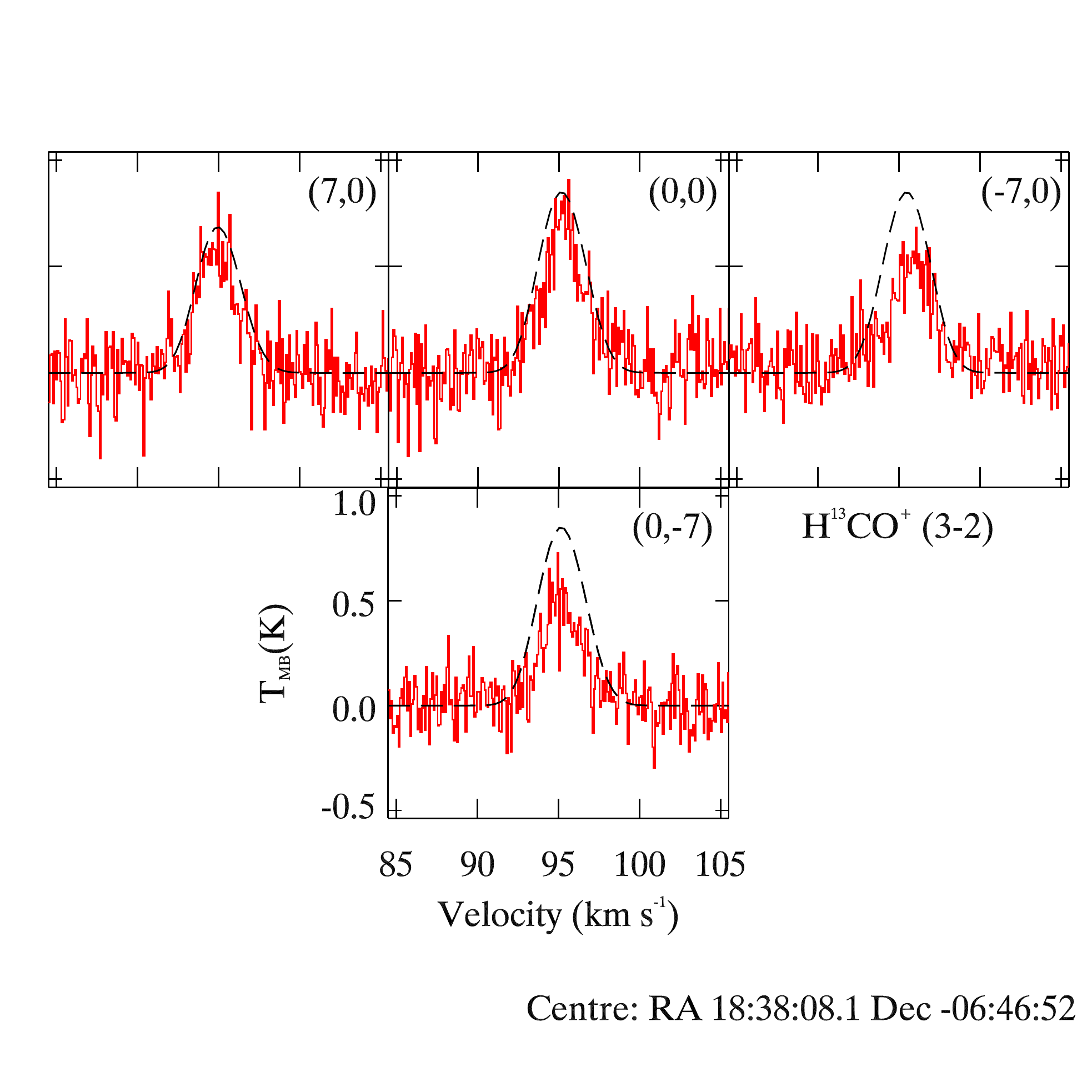}
    \caption{H$^{13}$CO$^{+}$ $(J = 3~$--$~2)$ line spectra taken with the JCMT. The frequency of this transition is 260.25 GHz and the spectral resolution is 0.09~km~s$^{-1}$. The line profile positions correspond to the centre of Fig.~\ref{Fig:rgb-spitzer} and each line profile is offset from each other by 10$^{\prime\prime}$.}
    \label{Fig:h13coplus-32}
\end{figure}

HCO$^{+}$ and H$^{13}$C O$^{+}$ were used to trace infalling
molecular gas. Fig.~\ref{Fig:hcoplus-10} ~\&~\ref{Fig:hcoplus-32}
show the line emission of two different rotational transitions
from the HCO$^{+}$ molecule. Both line profiles show a clear blue
asymmetric line shape which is characteristic of infalling gas.
The lines are fit with a model that includes infalling gas with a
velocity of 1.5~km~s$^{-1}$. Two rotational transitions of
H$^{13}$CO$^{+}$ are shown in Fig.~\ref{Fig:h13coplus-10}
~\&~\ref{Fig:h13coplus-32}. The line profiles are single peaked
though there may be a slight shoulder in the spectra of the
$(1~$--$~0)$ transition as the opacity of the radiation is higher
in the lower transitions. In contrast to HCO$^{+}$ the signature
of infalling gas is not seen in the line emission from
H$^{13}$CO$^{+}$ because it is not as abundant enough that the
optical thickness can get very high. Single peaked H$^{13}$CO$^+$
lines also effectively rule out the possibility that the double
peaked HCO$^+$ profiles could be caused by two cores.

Figure~\ref{Fig:h13coplus-10} \&~\ref{Fig:h13coplus-32} are
optically thin transitions that peak in intensity at the same
velocity as the $(J = 1~$--$~0)$ \& $(J = 2~$--$~1)$ transitions
of C$^{18}$O and the $(J = 1~$--$~0)$, $(J = 2~$--$~1)$ and $(J =
3~$--$~2)$ transitions of C$^{17}$O. The H$^{13}$CO$^{+}$ molecule
has an abundance significantly less than HCO$^{+}$ which effects
the observed line profile in two ways. Firstly even though it is
tracing infalling gas there is no self absorption and therefore a
double peaked, blue asymmetric spectral line is not observed.
Secondly the spectral line is predominately Gaussian and peaks at
the systemic velocity of the molecular cloud.

The velocity at which optically thin gas peaks in intensity (see
Fig.~\ref{Fig:h13coplus-32}) should coincide with an absorption
dip in line profiles from infalling optically thick gas. In
Fig.~\ref{Fig:hcoplus-10} \&~\ref{Fig:h13coplus-32} such a
characteristic line profile is seen suggesting that these lines
are greatly effected by infalling gas. In both species the infall
velocity of the gas is constant as a function of distance from the
centre of the cloud but a velocity shift similar to that detected
in the optically thin CO species
(Section~\ref{sect:optically-thin}) of 0.5~km~s$^{-1}$ is also
seen in these line profiles.

\begin{figure}
    \centering
    \includegraphics[width=6cm]{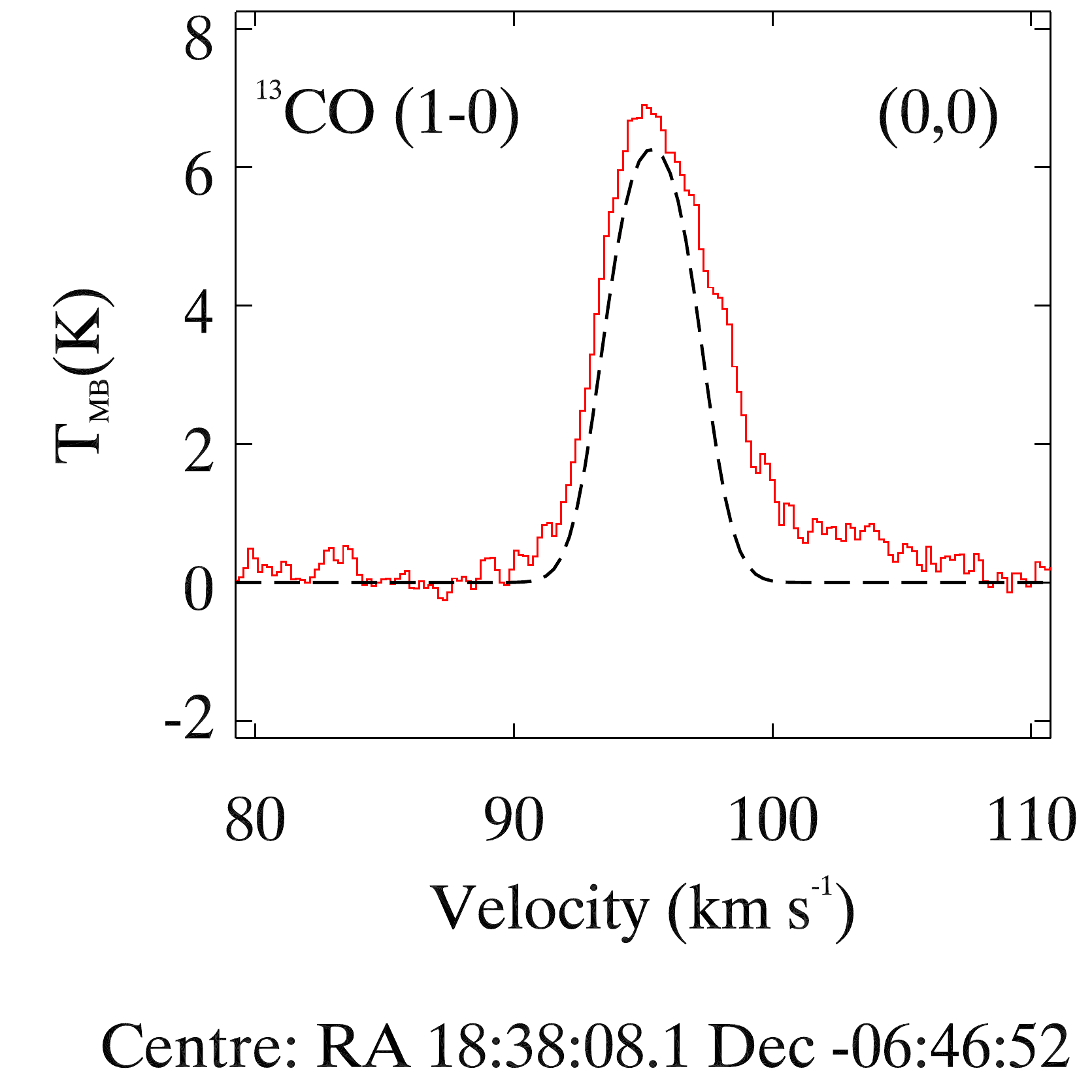}
    \caption{$^{13}$CO $(J = 1$--$~0)$ line spectra taken with the Mopra telescope. The frequency of this transition is 110.20 GHz and the spectral resolution is 0.23~km~s$^{-1}$. The line profile position corresponds to the centre of Fig.~\ref{Fig:rgb-spitzer}.}
    \label{Fig:13co-1-0}
\end{figure}

The line emission spectra from the $(J = 1~$--$~0)$ rotational
transition of $^{13}$CO is shown in Fig.~\ref{Fig:13co-1-0}. The
critical density of this transition is three orders of magnitude
less than both the HCO$^{+}$ \& H$^{13}$CO$^{+}$ lines and the
peak density of molecular hydrogen in this cloud. Therefore the
emission that is observed is originating from a different
dynamical region than HCO$^{+}$ or H$^{13}$CO$^{+}$. The line
shape in Fig.~\ref{Fig:13co-1-0} has a slight blue asymmetry which
indicates infalling gas. The asymmetry is not too severe and there
is no double peaked line shape which indicates this line is
tracing the slowest infalling gas that resides at the out-most
edge of the infall region.

\subsubsection{HCN as a dynamical tracer?}

\begin{figure}
    \centering
    \includegraphics[width=6cm]{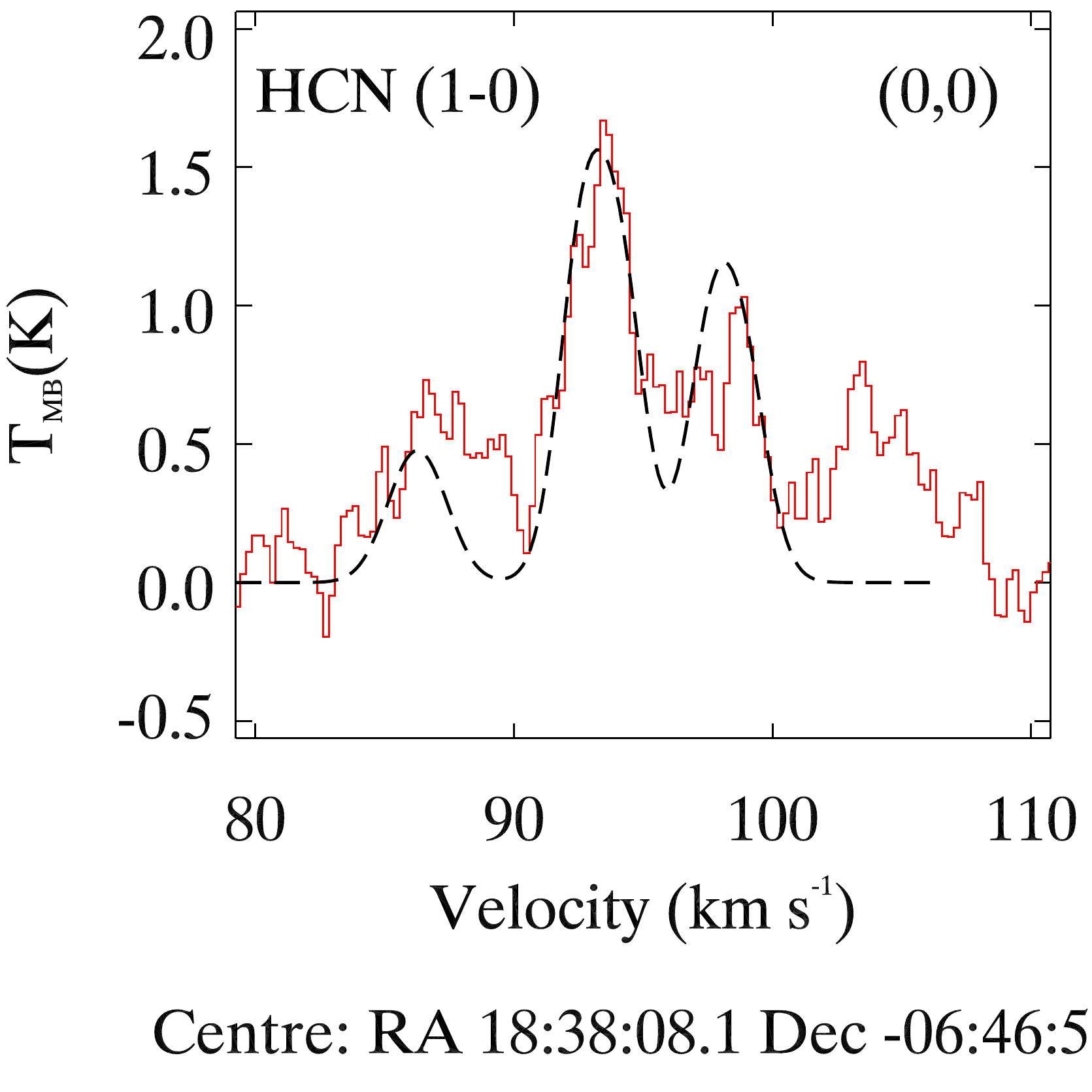}
    \caption{HCN $(J = 1~$--$~0)$ line spectra taken with the Mopra telescope. The frequency of this transition is 88.63 GHz and the spectral resolution is 0.23~km~s$^{-1}$. The line profile position corresponds to the centre of Fig.~\ref{Fig:rgb-spitzer}.}
    \label{Fig:hcn-10}
\end{figure}

\begin{figure}
    \centering
    \includegraphics[width=8cm]{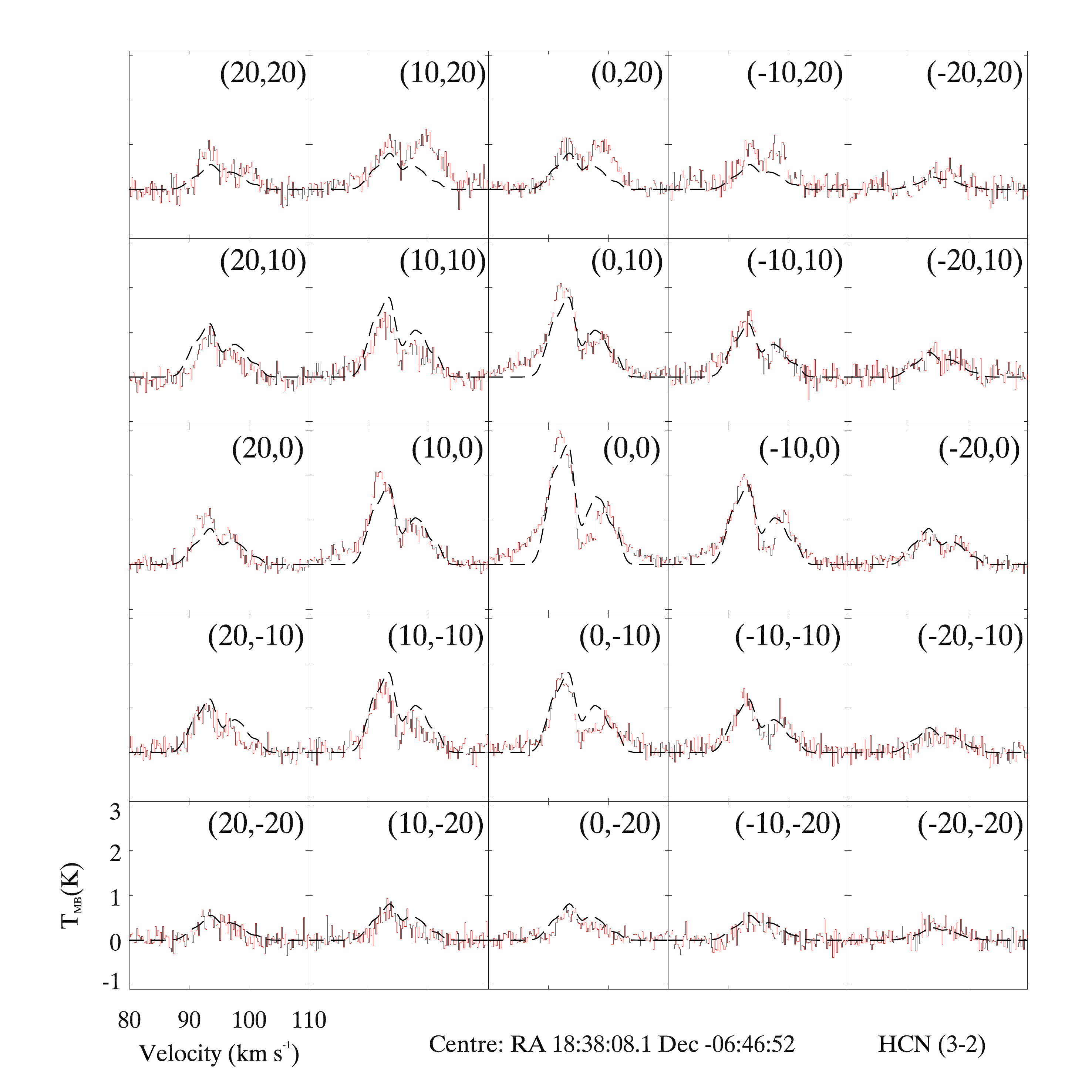}
    \caption{HCN $(J = 3~$--$~2)$ line spectra taken with the JCMT. The frequency of this transition is 265.88 GHz and the spectral resolution is 0.21~km~s$^{-1}$. The line profile positions correspond to the centre of Fig.~\ref{Fig:rgb-spitzer} and each line profile is offset from each other by 10$^{\prime\prime}$.}
    \label{Fig:hcn-32}
\end{figure}

The HCN line profiles are shown in Fig.~\ref{Fig:hcn-10} and
\ref{Fig:hcn-32} and show the $(J = 1~$--$~0)$~\&~$(J = 3~$--$~2)$
rotational transitions. The critical density $n_{\rm{crit}}$ of
these transitions are $2.6 \times 10^{6}$ and $5.2 \times 10^{7}$
cm$^{-3}$ which are both higher than the peak density of molecular
hydrogen in our model ($n_{\rm{H}_{2}}$ = 10$^{6}$ cm$^{-3}$).

When the density is larger than $n_{\rm{crit}}$ an excited
molecule will transfer its energy to a collision partner such as
H$_{2}$ and He before it has time to emit the radiation as a
photon. This collisional de-excitation is important for low
critical density species such as CO, however the HCN lines shown
in Fig.~\ref{Fig:hcn-10} and \ref{Fig:hcn-32} arise from gas that
is dominated by radiative de-excitation. This means the rotational
transitions that are observed here probe the gas in the dense
core. The $(J = 1~$--$~0)$ line profile may be effected to some
extent by collisional de-excitation as its critical density is
close to the peak density of molecular hydrogen.

The HCN emission profile contains hyperfine line emission which
complicates the observed line profile. This is unfortunate for
using HCN as a gas dynamical tracer species for distant massive
star forming regions because it does have other features, such as
a large critical density, that make it an ideal physical probe.
The drawback is the hyperfine structure of the rotational lines
due to the nuclear quadrupole moment of $^{14}$N.  At the low J
levels, the hyperfine components are well separated and at higher
J levels they are blended. The HCN $(J = 1~$--$~0)$ line for
example has hyperfine components spread over $\sim 13~{\rm
km~s^{-1}}$. A further complication, beyond the scope of this
work, is that HCN often exhibits poorly understood 'hyperfine
anomalies' where the individual hyperfine components are boosted
or suppressed far beyond what would be expected if they are
thermal equilibrium with each other
\citep{bains-09,guilloteau&baudry-81}.

The hyperfine structure is just resolved in our HCN $(J =
1~$--$~0)$ observations. It is not clear whether hyperfine
anomalies are present in this line data because the dynamics of
the source plus the usual opacity effects along the line of sight
lead to a highly confused line profile shape and somewhat poor
model fit. While the hyperfine structure is not resolved at HCN
$(J = 3~$--$~2)$ its underlying effects are apparent in the larger
observed line splitting due to infall of HCN $(J = 3~$--$~2)$ than
HCO$^{+}$ $(J = 3~$--$~2)$.

In general, it is a mistake to select HCN $(J = 3~$--$~2)$ as an
infall tracer for quantitative work simply on the basis that the
line splitting is sharper since in fact this is due to the
underlying hyperfine structure. In addition, the possible presence
of poorly understood HCN hyperfine anomalies means that this
molecule must be used with great caution as a dynamical tracer.
This point is being developed by Redman et al (2009, in prep).

\subsubsection{Molecular Outflow}

Jet-driven molecular outflows are detected alongside cores forming
massive stars \citep{shepherd&churchwell96a,shepherd05}, low mass
stars \citep{lada85,fukui89,reipbally01,cabrit03} and brown dwarfs
\citep{whelan-et-al07,phan-bao08,whelan09} which suggests they are
ubiquitous to the star formation process. Their role is to remove
angular momentum in the core that results from spin up due to
accreting gas. Outflowing gas tends to have broad wings of
emission indicating gas at a high velocity with respect to the
systemic velocity of the core. The optically thick lines $^{12}$CO
often have two distinct peaks of emission separated in velocity by
a few km s$^{-1}$. This is readily seen in observed line emission
profiles of the lower rotational transitions.

\begin{figure*}
    \centering
    \includegraphics[width=15cm]{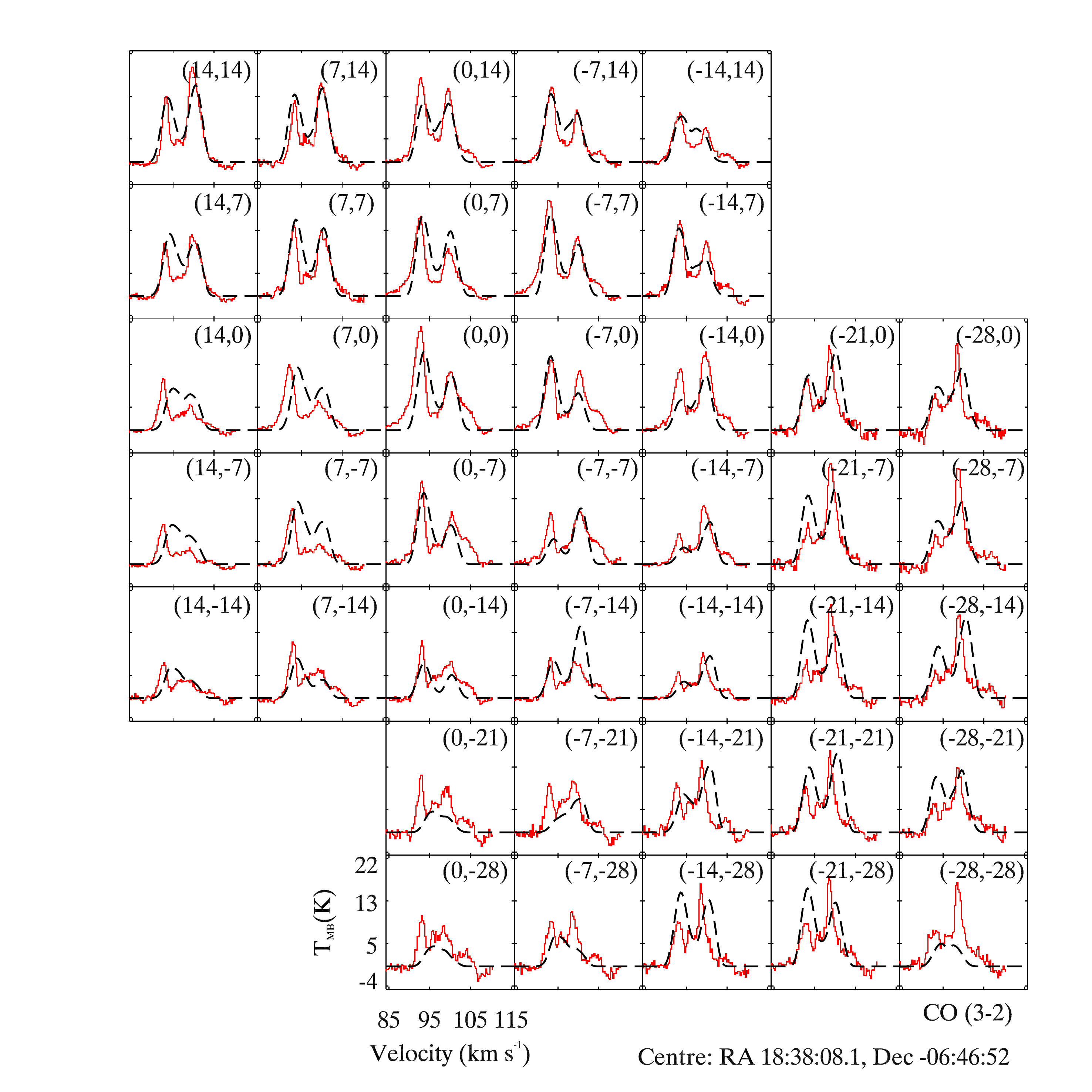}
    \caption{$^{12}$CO $(J = 3~$--$~2)$ line spectra taken with the JCMT. The frequency of this transition is 345.79 GHz and the spectral resolution is 0.27~km~s$^{-1}$. The line profile positions correspond to the centre of Fig.~\ref{Fig:rgb-spitzer} and each line profile is offset from each other by 7$^{\prime\prime}$.}
    \label{Fig:model-12co}
\end{figure*}

\begin{figure}
    \centering
    \includegraphics[width=8cm]{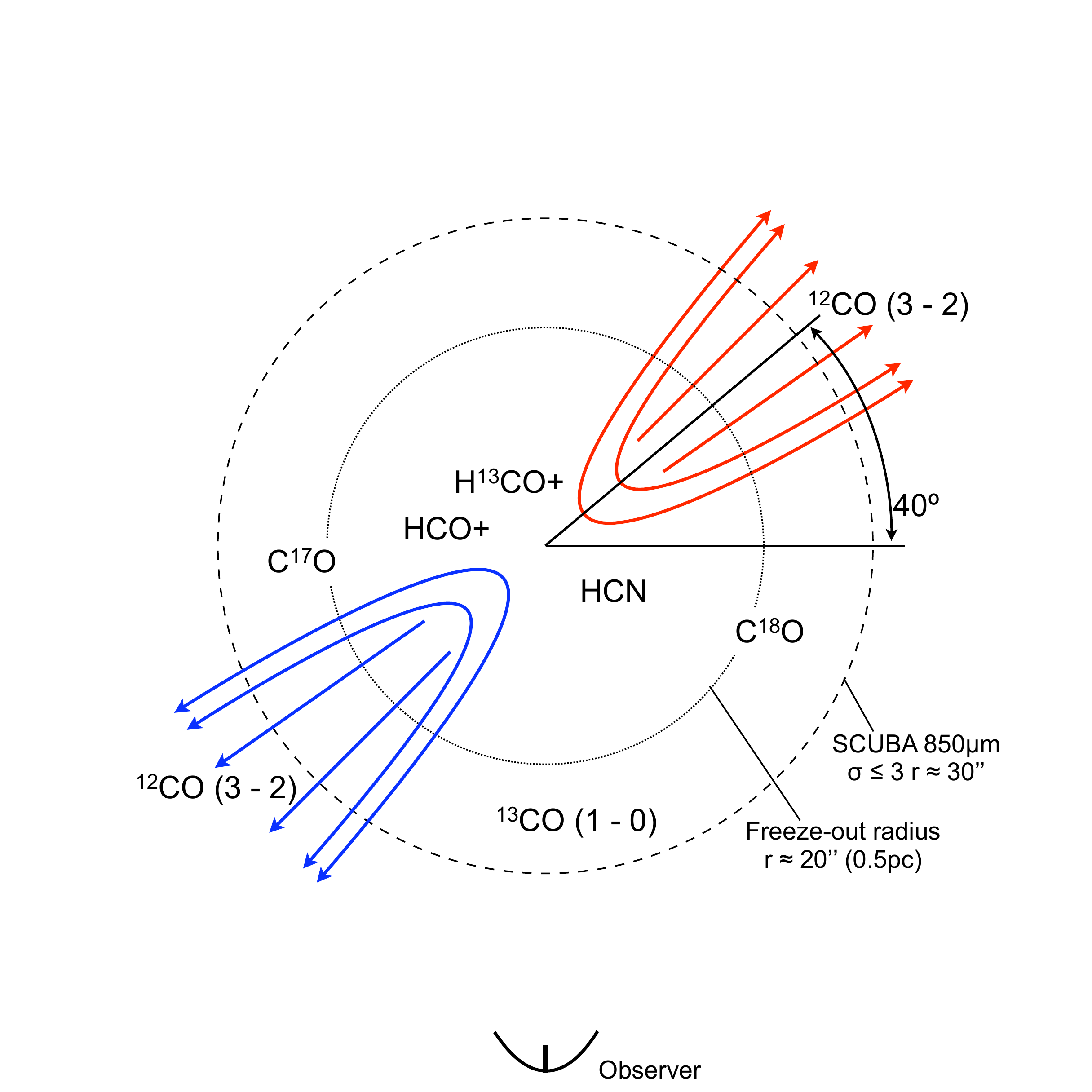}
    \caption{A schematic representation of the regions in JCMT\,18354S that are traced by each molecular line. The orientation of the outflow in the plane of the sky and the lobes of red and blue shifted emission are indicated with different colours.}
    \label{Fig:schematic}
\end{figure}

Figure~\ref{Fig:model-12co} shows the CO $(J = 3~$--$~2)$
rotational transition line emission over a 50$^{\prime\prime}$
$\times$ 50$^{\prime\prime}$ area. The lines profiles are clearly
double peaked and asymmetric. The asymmetry of the line shape
switches from blue-asymmetric at (0,0) to red-asymmetric at
(-21,-14). Such switches in peaks of intensity are also seen in
the outflow lobes in low mass star formation. The observed line
profiles were fit with a model that includes all the physical
parameters constrained from the observations presented earlier.
$^{12}$CO is the second most abundant molecule in molecular
clouds. This makes it a good candidate for tracing the molecular
outflow since an outflow occupies a small fraction of the volume
of a cloud. Our best fit model indicates the emission is coming
from a bipolar structure. The orientation of the outflow with
respect to the line of sight creates the double peaks of the
spectra in Fig.~\ref{Fig:model-12co}. The double peaks represent
emission that is blue-shifted and red-shifted because the
outflowing gas is pointed towards and away from the observer
respectively \citep{rawlings.et.al04}. The asymmetry in the
intensity changes with the orientation of the outflow axis to the
line of sight. An angle of $30^{\circ} \rightarrow 40^{\circ}$ for
the outflow axis to the line of sight gave the best fit to the
observed line profiles. It should be noted that the peak of
$^{12}$CO $(J = 3~$--$~2)$ emission coincides with the peak of the
dust continuum emission as seen in Fig.~\ref{Fig:rgb-spitzer}.
This effect is due to a combination of the orientation of the
outflow axis to the line of sight and the large beam size of the
line emission observations. A schematic of the structure of JCMT
18354S in Fig.~\ref{Fig:schematic} and it indicates how the
orientation of the outflow axis will result in the gas from the
outflow lobes coinciding in the line of sight with the centre of
the cloud.

The $^{12}$CO line profiles are better fit in the outlying regions
rather than in the centre of the core. This is caused by a rapidly
changing temperature and morphology in the centre. In contrast the
gas that is farther out in the lobes of the outflow comes from a
region where conditions are much more stable. The result is that
gas originating in the lobes is easier to model whereas in the
centre where the gas is changing rapidly and localised in a small
region, it is more difficult.

The presence of outflowing gas is a good indicator of star
formation activity in low mass protostellar cores and the
modelling here demonstrates the major effect on the $^{12}$CO $(J
= 3~$--$~2)$ line emission observed in JCMT\,18354S. The
blue-shifted outflow emission coincides with the peak in emission
of the IRAC image in Fig.~\ref{Fig:rgb-spitzer}. There is no IRAC
emission on the red-shifted side because it is obscured by dense
gas in the line of side. In addition, archival SiO $(J =
6~$--$~5)$ and $(J = 2~$--$~1)$ emission line profiles (which are
not displayed or modelled, since SiO is not yet fully incorporated
into mollie) peak at the same location as the IRAC emission. SiO
emission normally indicates recently shocked gas and in low mass
cores it is shown to trace the molecular outflow
\citep{santiago-garcia-et-al08}. We therefore consider the SiO and
IRAC emission to be tracing the molecular outflow.

\section{Discussion}
\label{discussion}

The model described in the previous section is of a massive
protostellar core undergoing infall, outflow, rotation and
depletion. The best fit physical parameters for each molecular
species for the model are shown in Table~\ref{lines}. Further
improvement to the fits, particularly the outflow wings line
profiles may be obtained by introducing a gradient in the
turbulent velocity, allowing for velocity structure within the
outflow and incorporating a chemical model to allow for more
realistic variations in the species abundances.

So that the properties of JCMT\,18354S can be compared to those
prevalent in low mass star formation, the results of a modelling
exercise for a typical nearby low mass star forming core, L483
\citep{carolan-et-al-08} are also presented in Table~\ref{lines}.
The central density and radius (and thus overall mass) of the cold
massive core is higher as would be expected. However many of the
parameters are strikingly similar between the two cores. The
physical properties of the cloud gas such as the temperature,
degree of depletion and molecular abundances of the low and high
mass cores are comparable. The global behaviour of the massive
star forming system is also similar to that of low mass star
formation: infall, rotation and outflow are all implied by the
observations and modelling. These dynamical processes act on
larger physical scales and support the interpretation of massive
star formation as a scaled up version of low mass star formation,
at least at this evolutionary stage.

The most significant differences between the two classes of object
are in the infall and turbulent velocities, both of which are an
order of magnitude greater than gas in low mass cores. This
difference is very significant because in cold massive cores, such
velocities are highly supersonic compared to the isothermal sound
speeds. The large velocity in both the infalling and turbulent gas
is ultimately due to the large gravitational potential well at the
centre of the molecular cloud. The dynamics of the cold massive
core are thus dominated by supersonic velocities.

The supersonic turbulence could be interpreted literally as just
that: very energetic turbulence in one large gas cloud. Supersonic
turbulence will undergo collisions of gas streams and the
formation of localised eddies. Shocks must result since the
motions are highly supersonic which will lead to localised
dissipation of energy and then may lead to the formation of dense
clumps. These clumps may either become self gravitating or more
transient structures that then quickly dissipate
\citep{garrod.et.al05}. The observations presented here do not
have the spatial resolution to resolve the small scale structure
to test the clumpy cloud scenario discussed in \cite{keto&wood06}.
However, it is possible that clumps existed prior to the collapse
of the cloud and that the observed supersonic velocities are due
to the net effect of individual clumps moving around the
gravitational well. It is worth noting that in low mass star
formation, very long lived gravitationally stable starless cores
are observed i.e. in quiescent environments at least, undisturbed
clumps can persist without either forming stars or dissipating
\citep{redman06}. Later in the massive star formation process,
such clumps may effect the development of the UCH{\sc ii} region
\citep{redman.et.al98b} but presumably they are the objects that
collapse to form the cluster of hundreds or thousands of low mass
stars that form alongside massive stars. A radiative transfer and
dynamical model of a clumpy accretion flow will be developed
further in future work.

\section{Conclusions}
\label{conclusions}

The high mass protostellar core JCMT\,18354S was observed in
sixteen different molecular line transitions at multiple spatial
positions and in dust continuum emission. The presence of
so-called 'green fuzzy' 4.5$\mu$m emission \citep{chambers09}, SiO
emission and broad absorption lines in JCMT\,18354S are all
consistent with the characteristics of a deeply embedded massive
star forming core \citep{beuther07}. However, no 24$\mu$m
emission, which would come from warm dust in the vicinity of the
protostellar object, is detected.

The entire line data-set was self-consistently modelled using {\sc
mollie}, a 3D molecular line transfer code. The large data-set
tightly constrained the source model to be that of a rotating,
depleted, infalling envelope collapsing onto a central
protostellar source that has evolved sufficiently to generate a
molecular outflow. Comparing these properties to those prevalent
in low mass star forming core shows very clearly that JCMT\,18354S
is a scaled up version of a low mass star forming core. The major
difference is the highly supersonic internal velocities in this
massive core compared to those of low mass cores. This may lead to
the formation of gas clumps which in turn become a population of
low mass stars.

\section*{Acknowledgements}
We thank the referee for a prompt and constructive report that led
to an improved paper. This research was supported through a
Science Foundation Ireland (SFI) Research Frontiers award. PBC
received support from the Cosmo-Grid project, funded by the
Program for Research in Third Level Institutions under the Irish
National Development Plan and with assistance from the European
Regional Development Fund. RML was supported by an Irish Research
Council for Science Engineering and Technology studentship. We
also acknowledge SFI/Higher Education Authority Irish Centre for
High-End Computing (ICHEC) for the provision of computational
facilities and support.

\bibliographystyle{mn2e}
\bibliography{paper-i18354s}
\end{document}